\documentclass[aip,amsmath,amssymb,reprint,]{revtex4-1}
\usepackage{graphicx}
\usepackage{color}
\usepackage{dcolumn}
\usepackage{bm}
\usepackage[utf8]{inputenc}
\usepackage[T1]{fontenc}
\usepackage{mathptmx}
\usepackage{etoolbox}
\usepackage{mathtools}
\usepackage{hyperref}
\makeatletter
\def\@email#1#2{%
 \endgroup
 \patchcmd{\titleblock@produce}
  {\frontmatter@RRAPformat}
  {\frontmatter@RRAPformat{\produce@RRAP{*#1\href{mailto:#2}{#2}}}\frontmatter@RRAPformat}
  {}{}
}%

\makeatother
\begin{document}

\title{Ground state degeneracy on torus in a family of $\mathbb{Z}_N$ toric code}

\author{Haruki Watanabe}
\altaffiliation{Department of Applied Physics, The University of Tokyo.}

\author{Meng Cheng}
\altaffiliation{Department of Physics, Yale University.}

\author{Yohei Fuji}
\altaffiliation{Department of Applied Physics, The University of Tokyo.}

\date{\today}

\begin{abstract}
Topologically ordered phases in $2+1$ dimensions are generally characterized by three mutually-related features:
fractionalized (anyonic) excitations, 
topological entanglement entropy, and
robust ground state degeneracy that does not require symmetry protection or spontaneous symmetry breaking.
Such degeneracy is known as topological degeneracy and usually can be seen under the periodic boundary condition regardless of the choice of the system size $L_1$ and $L_2$ in each direction. In this work we introduce a family of extensions of the Kitaev toric code to $N$ level spins ($N\geq2$).
The model realizes topologically ordered phases or symmetry-protected topological phases depending on parameters in the model.
The most remarkable feature of the topologically ordered phases is that the ground state may be unique, depending on $L_1$ and $L_2$, despite that the translation symmetry of the model remains unbroken. Nonetheless, the topological entanglement entropy takes the nontrivial value. We argue that this behavior originates from the nontrivial action of translations permuting anyon species. 
\end{abstract}

\maketitle

\section{Introduction}
In the studies of many-body systems, one is often interested in the properties of ground states and low energy excitations. Ground state degeneracy  that does not originate from spontaneous symmetry breaking or fine-tuning of parameters is called topological degeneracy~\cite{WenBook,WenZoo,XieBook}. Such a degeneracy is robust against any local perturbations including symmetry-breaking ones.
In two dimensions, the order of topological degeneracy $N_{\text{deg}}$ depends on the genus $g$ of the manifold on which the system is defined. 

In topologically ordered phases with U(1) symmetry (e.g., fractional quantum Hall systems), $N_{\text{deg}}\geq q^g$ when the filling is $\nu=1/q$.
This degeneracy can be proven by a flux-threading type argument~\cite{OshikawaSenthil} assuming the appearance of fractional excitations with U(1) charge $1/q$.  More generally,  there usually exist closed string operators that describe processes of creating a pair of anyonic excitations, dragging them apart, and pair-annihilating them again after forming a non-contractible loop. These loops commute with the Hamiltonian but not among them. Non-commutativity of loop operators implies the topological degeneracy. {In particular, the topological degeneracy $N_\text{deg}$ on a torus ($g=1$) is often equal to the number of distinct anyonic excitations. The topological degeneracy is also tied with the topological entanglement entropy \cite{KitaevPreskill, LevinWen}, which is given by $S_{\mathrm{topo}}=-\log\mathcal{D}$ where $\mathcal{D}$ is the total quantum dimension; $\mathcal{D}^2$ is nothing but the number of distinct anyonic excitations for Abelian topological order.  Therefore, it is often stated that the ground state degeneracy on torus, anyonic excitations, and topological entanglement entropy appear all at the same time}. 

In this work, we introduce a family of extensions of the Kitaev toric code \cite{doi:10.1063/1.1499754,KITAEV20032} to $N$-level spins ($N=2,3,4,\cdots$), which contains an integer parameter $a$ ($1\leq a\leq N$).
The original model corresponds to the $(N,a)=(2,1)$ case. The model describes topologically ordered phases when $a$ is not a multiple of $\mathrm{rad}(N)$ (the radical of $N$; see Sec.~\ref{classification}) and phases with no topological order when $a$ is a multiple of $\mathrm{rad}(N)$. In particular, when $N$ and $a$ are coprime, these phases are characterized by the topological entanglement entropy $S_{\mathrm{topo}}=-\log N$, independent of system size $L_1,L_2$.

For a generic integer $N$, the case of $a=1$ is the standard $\mathbb{Z}_N$ toric code~\cite{KITAEV20032} discussed widely, for example, in Refs.~\onlinecite{BullockBrennen,PhysRevB.94.075151,SlagleKim,Vijay2017}, which shows topological degeneracy $N_{\mathrm{deg}}=N^2$ regardless of the choice of $L_1$ and $L_2$.
The case of $a=N-1$ ($N\geq3$) was discussed in Refs.~\onlinecite{Schulz2012,MengZN,PhysRevB.96.245122,FujiZN}, although the ground state degeneracy on torus was not fully investigated. When $N$ is odd and $a=N-1$, we find that the topological degeneracy occurs only when both $L_1$ and $L_2$ are even:
\begin{align}
N_{\mathrm{deg}}=
\begin{cases}
N^2 &  \text{(Both $L_1$ and $L_2$ are even)}\\
1 & \text{(otherwise)}
\end{cases}.\label{most2}
\end{align}
The most striking situation of our model arises when $N$ is a prime number and $a$ is a primitive root modulo $N$ (see Sec.~\ref{mo}). In this case, $N_{\mathrm{deg}}$ is given by (see Sec.~\ref{MNa} for the proof)
\begin{align}
N_{\mathrm{deg}}=
\begin{cases}
N^2 &  \text{(Both $L_1$ and $L_2$ are multiples of $N-1$)}\\
1 & \text{(otherwise)}
\end{cases}.\label{most1}
\end{align}
This means that the minimum system size to observe the degeneracy is $L_1=L_2=N-1$, for which the Hilbert space dimension is $N^{2(N-1)^2}$ (for example, $11^{200}$ for $N=11$, for which $a=2,6,7,8$ are the primitive roots). It is thus nearly impossible to see the topological degeneracy for a large $N$ in any numerical studies.  Therefore, the uniqueness of the ground state for a sequence of $L_i$ cannot be used as a proof of the absence of topological order, although the converse might still be the case: topological degeneracy $N_{\mathrm{deg}}>1$ in a sequence of $L_i$ implies a nontrivial topological order. Note that, if an  \emph{open} boundary condition is assumed instead of the periodic one, a unique ground state can be realized even in the original $\mathbb{Z}_2$ toric code due to the absence of any Wilson loops or constraints among stabilizers.

There is a more famous example, called Wen's plaquette model~\cite{PhysRevLett.90.016803}, in which topological degeneracy depends on the system size. 
There are also more recent examples of this type behavior~\cite{PhysRevB.100.125150,PhysRevB.105.045128,PhysRevB.106.045145,2207.00409,2204.01279}.
However, in these examples,  the ground state degeneracy on torus is at least two.  
Our example demonstrates that there are even cases where the ground state is unique and excitations are all gapped in a sequence of $L_i$, despite their nontrivial topological order.  It is interesting to contrast with a known theorem about topological quantum field theory (TQFT),  according to which the phase is invertible (i.e., no topological order) if $N_{\mathrm{deg}}=1$ on torus (and technically, on sphere as well)~\cite{Schommer}. Our example shows that the relation between lattice models and corresponding effective field theories can be quite subtle. We will also show that the degeneracy can be understood in terms of the TQFT if the finite-size torus in the lattice system is viewed as a torus in continuum but with symmetry defect lines (or twisted boundary conditions) corresponding to the translation symmetry action in the low-energy theory.

The rest of this work is organized as follows. 
We summarize the definition and basic properties of our model in Sec.~\ref{sec:model}.
We review basic mathematical facts in number theory in Sec.~\ref{sec:math}. Overall properties of our model for a given integers $N$ and $a$ are summarized in Sec.~\ref{classification}.
Then the ground state degeneracy of the model in topologically ordered phases is studied in Sec.~\ref{sec:case1}. 
The relation of our  model to the standard $\mathbb{Z}_N$ toric code model is clarified in Sec.~\ref{sec:relation}.
Topological properties such as the topological entanglement entropy and anyon statistics in our model are discussed in Sec.~\ref{sec:topology}. 
Finally, we study the cases with no topological order in Sec.~\ref{sec:case2}.
We then conclude in Sec.~\ref{sec:discussion}.

\section{Definition of model}
\label{sec:model}
In this section, we explain the definition and the basic properties of $\mathbb{Z}_N$ toric code. Throughout this work, $N$ is an integer greater than $1$.

\begin{figure*}
\includegraphics[width=\textwidth]{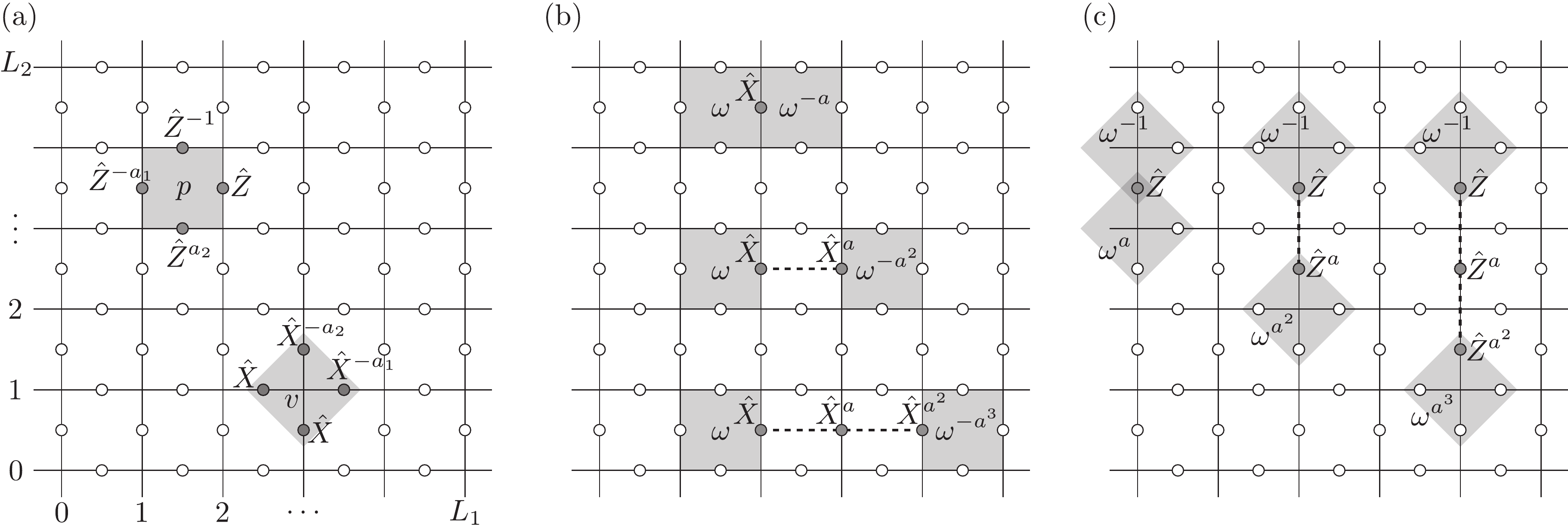}
\caption{\label{fig1} 
(a) The definition of a vertex operator $\hat{A}_v$ and a plaquette operator $\hat{B}_p$. 
(b) Pairs of magnetic excitations created by $\hat{X}_{\bm{r}}^{a^\ell}$.
(c) Pairs of electric excitations created by $\hat{Z}_{\bm{r}}^{a^\ell}$.}
\end{figure*}

\subsection{Lattice of $N$-level spins}
In our model, an $N$-level spin is placed on each link of square lattice. See Fig.~\ref{fig1} for the illustration.  The action of operators $\hat{X}_{\bm{r}}$ and $\hat{Z}_{\bm{r}}$ on the $N$-level spin at $\bm{r}$ is represented by $N$-dimensional unitary matrices
\begin{align}
&X\coloneqq\begin{pmatrix}
&&&&1\\
1&&&&\\
&1&&&\\
&&\ddots&&\\
&&&1&\\
\end{pmatrix},\label{defX}\\
&Z\coloneqq\begin{pmatrix}
1&&&&\\
&\omega&&&\\
&&\omega^2&&\\
&&&\ddots&\\
&&&&\omega^{N-1}\\
\end{pmatrix}\label{defZ},\\
&\omega \coloneqq e^{\frac{2\pi i}{N}},
\end{align}
which generalize the Pauli matrices. All matrix elements left blank are zero. They satisfy
\begin{align}
&Z^N=X^N=1,\\
&ZX=\omega\, XZ=
\begin{pmatrix}
&&&&1\\
\omega&&&&\\
&\omega^2&&&\\
&&\ddots&&\\
&&&\omega^{N-1}&\\
\end{pmatrix}.
\end{align}
Operators for different spins commute, so that
\begin{align}
&\hat{Z}_{\bm{r}}^N=\hat{X}_{\bm{r}}^N=1,\label{ZXN}\\
&\hat{Z}_{\bm{r}}\hat{X}_{\bm{r}'}=\omega^{\delta_{\bm{r},\bm{r}'}}\, \hat{X}_{\bm{r}'}\hat{Z}_{\bm{r}}.\label{ZXwXZ}
\end{align}
Note that, when $N\geq3$, matrices $X$ and $Z$ are not Hermitian and $\hat{X}_{\bm{r}}^\dagger\neq\hat{X}_{\bm{r}}$ and $\hat{Z}_{\bm{r}}^\dagger\neq\hat{Z}_{\bm{r}}$.

The role of $\hat{X}_{\bm{r}}$ and $\hat{Z}_{\bm{r}}$ can be interchanged by the global unitary transformation $\hat{U}_Y$, whose action on each spin is represented by
\begin{align}
&U_Y\coloneqq\frac{1}{\sqrt{N}}\begin{pmatrix}
1&1&1&\cdots&1\\
1&\omega&\omega^2&\cdots&\omega^{N-1}\\
1&\omega^2&\omega^4&\cdots&\omega^{2(N-1)}\\
\vdots&\vdots&\vdots&\ddots&\vdots\\
1&\omega^{N-1}&\omega^{2(N-1)}&\cdots&\omega^{(N-1)^2}\\
\end{pmatrix}.
\end{align}
We have $U_YXU_Y^\dagger =Z$ and $U_YZU_Y^\dagger =X^\dagger$.

The positions of spins on the horizontal and vertical links are set to $\bm{r}=(m_1+\frac{1}{2},m_2)$ and $(m_1,m_2+\frac{1}{2})$, respectively, where $m_i=0,1,\cdots,L_i-1$ ($i=1,2$) and $L_i$ is a positive integer:
\begin{align}
&\Lambda\coloneqq\{(m_1,m_2+\tfrac{1}{2}),(m_1+\tfrac{1}{2},m_2)\,|\,m_i=0,1,\cdots,L_i-1\}.
\end{align}
We impose the periodic boundary condition and identify $\hat{X}_{\bm{r}+(n_1L_1,n_2L_2)}$ with $\hat{X}_{\bm{r}}$ and $\hat{Z}_{\bm{r}+(n_1L_1,n_2L_2)}$ with $\hat{Z}_{\bm{r}}$ for any $n_1,n_2\in\mathbb{Z}$ and $\bm{r}\in\Lambda$.
The sets of vertices $\mathcal{V}$ and plaquettes $\mathcal{P}$ are given by
\begin{align}
&\mathcal{V}\coloneqq\{(m_1,m_2)\,|\,m_i=0,1,\cdots,L_i-1\},\\
&\mathcal{P}\coloneqq\{(m_1+\tfrac{1}{2},m_2+\tfrac{1}{2})\,|\,m_i=0,1,\cdots,L_i-1\}.
\end{align}
The total number of spins in the system is $2L_1L_2$ and the dimension of the Hilbert space is $N^{2L_1L_2}$.

\subsection{Hamiltonian and stabilizers}
The Hamiltonian of the model reads as
\begin{align}
\hat{H}\coloneqq-\sum_{v\in\mathcal{V}}\frac{1}{2}(\hat{A}_v+\text{h.c.})-\sum_{p\in\mathcal{P}}\frac{1}{2}(\hat{B}_p+\text{h.c.}),\label{defH}
\end{align}
which is invariant under translation $\hat{T}_i$ ($i=1,2$), defined by
\begin{align}
\hat{T}_i\hat{X}_{\bm{r}}\hat{T}_i^{\dagger}=\hat{X}_{\bm{r}+\bm{e}_i},\quad \hat{T}_i\hat{Z}_{\bm{r}}\hat{T}_i^{\dagger}=\hat{Z}_{\bm{r}+\bm{e}_i}.\label{translation}
\end{align}
As illustrated in Fig.~\ref{fig1} (a), a vertex operator $\hat{A}_v$ ($v\in \mathcal{V}$) is defined by
\begin{align}
\hat{A}_{(m_1,m_2)}&\coloneqq\hat{X} \overset{\displaystyle \hat{X}^{-a_2}}{\underset{\displaystyle \hat{X}}{\displaystyle +}} \hat{X}^{-a_1}\notag\\
&=\hat{X}_{(m_1+\frac{1}{2},m_2)}^{-a_1}\hat{X}_{(m_1,m_2+\frac{1}{2})}^{-a_2}\hat{X}_{(m_1-\frac{1}{2},m_2)}\hat{X}_{(m_1,m_2-\frac{1}{2})}\label{Av}
\end{align}
and a plaquette operator $\hat{B}_p$ ($p\in \mathcal{P}$) is 
\begin{align}
\hat{B}_{(m_1+\frac{1}{2},m_2+\frac{1}{2})}&\coloneqq\hat{Z}^{-a_1} \overset{\displaystyle \hat{Z}^{-1}}{\underset{\displaystyle \hat{Z}^{a_2}}{\displaystyle \Box}} \hat{Z}\notag\\
&=\hat{Z}_{(m_1+1,m_2+\frac{1}{2})}\hat{Z}_{(m_1+\frac{1}{2},m_2+1)}^{-1}\hat{Z}_{(m_1,m_2+\frac{1}{2})}^{-a_1}\hat{Z}_{(m_1+\frac{1}{2},m_2)}^{a_2}.\label{Bp}
\end{align}
Integers $a_1$ and $a_2$ ($1\leq a_1,a_2\leq N$) are important parameters of this model. It is easy to verify that $\hat{A}_v$'s ($v\in \mathcal{V}$) and $\hat{B}_p$'s ($p\in \mathcal{P}$) all commute with each other regardless of $a_1$ and $a_2$.  For brevity, we set $a_1=a_2=a$ in the following, but $a_1\neq a_2$ cases can be treated in the same way. 

The eigenstates of the Hamiltonian can be chosen as simultaneous eigenstates of all $\hat{A}_v$'s ($v\in \mathcal{V}$) and $\hat{B}_p$'s ($p\in \mathcal{P}$). Since 
\begin{align}
\hat{A}_v^N=\hat{B}_p^N=1,
\end{align}
eigenvalues of operators $\hat{A}_v$ and $\hat{B}_p$ are $N$-fold, $1,\omega,\cdots,\omega^{N-1}$.

\subsection{A ground state}
\label{sec:gs}
A ground state of the Hamiltonian $\hat{H}$ can be constructed explicitly following the discussion for the original toric code, for example, in Ref.~\onlinecite{TasakiBook}.
Let $|\phi_0\rangle$ be the ``ferromagnetic'' product state, satisfying 
\begin{align}
&\hat{Z}_{\bm{r}}|\phi_0\rangle=|\phi_0\rangle\quad (\forall \bm{r}\in\Lambda),\\
&\hat{T}_i|\phi_0\rangle=|\phi_0\rangle.
\end{align}
It has the eigenvalue $+1$ for all plaquette operators:
\begin{align}
\hat{B}_p|\phi_0\rangle=|\phi_0\rangle\quad (\forall p\in \mathcal{P}).
\end{align}

Now we introduce a projection operator
\begin{align}
&\hat{P}\coloneqq \frac{1}{N^{L_1L_2}}\prod_{v\in\mathcal{V}}\sum_{\ell=0}^{N-1}\hat{A}_v^\ell,
\end{align}
which satisfies
\begin{align}
&\hat{P}^2=\hat{P},\\
&\hat{A}_{v}\hat{P}=\hat{P}\hat{A}_{v}=\hat{P},\\
&\hat{B}_{p}\hat{P}=\hat{P}\hat{B}_{p},\\
&\hat{T}_{i}\hat{P}=\hat{P}\hat{T}_{i}.
\end{align}
Then the state 
\begin{align}
&|\Phi_0\rangle\coloneqq\sqrt{\frac{N^{L_1L_2}}{N_C}}\hat{P}|\phi_0\rangle,\label{gs}
\end{align}
satisfies both
\begin{align}
&\hat{A}_v|\Phi_0\rangle=|\Phi_0\rangle\quad (\forall v\in \mathcal{V}),\label{AvPhi}\\
&\hat{B}_p|\Phi_0\rangle=|\Phi_0\rangle\quad (\forall p\in \mathcal{P}),\label{BpPhi}
\end{align}
suggesting that $|\Phi_0\rangle$ is a ground state with the energy eigenvalue $E_{\mathrm{GS}}=-2L_1L_2$.  
Here, $N_C>0$ is the normalization factor given by
\begin{align}
N_C=\langle\phi_0|\prod_{v\in\mathcal{V}}\sum_{\ell=0}^{N-1}\hat{A}_v^\ell|\phi_0\rangle=\sum_{\{\ell_v\}}\langle\phi_0|\prod_{v\in\mathcal{V}}\hat{A}_v^{\ell_v}|\phi_0\rangle.\label{Nc}
\end{align}
It can be shown that $N_C$ counts the number of global constraints among the vertex operators of the form 
\begin{align}
\prod_{v\in\mathcal{V}}\hat{A}_v^{\ell_v}=1\quad(0\leq \ell_v\leq N-1),\label{constraint}
\end{align}
and there is an equal number of constraints among the plaquette operators
\begin{align}
\prod_{p\in\mathcal{P}}\hat{B}_p^{\ell_p}=1\quad(0\leq \ell_p\leq N-1).\label{constraint2}
\end{align}
In our model, the total number of vertex operators and plaquette operators, $2L_1L_2$, coincides with the total number of spins in the system. Hence, there would be no ground state degeneracy if all stabilizers were independent. Indeed, as we demonstrate in Sec.~\ref{sec:case1},  the number of constraints $N_C$ is related to the ground state degeneracy as $N_{\mathrm{deg}}=N_C^2$.

The state $|\Phi_0\rangle$ is translation invariant
\begin{align}
\hat{T}_i|\Phi_0\rangle=|\Phi_0\rangle,
\end{align}
because $|\phi_0\rangle$ is translation invariant and $\hat{P}$ commutes with $\hat{T}_i$. As we show below, the model has a nonzero excitation gap. Furthermore, all correlation functions of $\hat{X}_{\bm{r}}$ and $\hat{Z}_{\bm{r}}$ are short-ranged. For example,
\begin{align}
\langle\Phi_0|\hat{X}_{\bm{r}}^\dagger \hat{X}_{\bm{r}'}|\Phi_0\rangle=\langle\Phi_0|\hat{Z}_{\bm{r}}^\dagger \hat{Z}_{\bm{r}'}|\Phi_0\rangle=\delta_{\bm{r},\bm{r}'}.
\end{align}
These observations imply the absence of translation symmetry breaking.

\subsection{Quasiparticle excitations}
\label{local}
We introduce (open) string operators $\hat{X}_{p,p'}^{(i)}$ ($i=1,2$) by
\begin{align}
\hat{X}_{(m_1-\frac{1}{2},m_2+\frac{1}{2}),(m_1'+\frac{1}{2},m_2+\frac{1}{2})}^{(1)}&\coloneqq\prod_{\ell=0}^{m_1'-m_1}\hat{X}_{(m_1+\ell,m_2+\frac{1}{2})}^{a^\ell},\label{m1}\\
\hat{X}_{(m_1+\frac{1}{2},m_2-\frac{1}{2}),(m_1+\frac{1}{2},m_2'+\frac{1}{2})}^{(2)}&\coloneqq\prod_{\ell=0}^{m_2'-m_2}\hat{X}_{(m_1+\frac{1}{2},m_2+\ell)}^{a^\ell}\label{m2}
\end{align}
and $\hat{Z}_{v,v'}^{(i)}$ ($i=1,2$) by
\begin{align}
\hat{Z}_{(m_1,m_2),(m_1'+1,m_2)}^{(1)}&\coloneqq\prod_{\ell=0}^{m_1'-m_1}\hat{Z}_{(m_1+\ell+\frac{1}{2},m_2)}^{a^{m_1'-m_1-\ell}},\label{e1}\\
\hat{Z}_{(m_1,m_2),(m_1,m_2'+1)}^{(2)}&\coloneqq\prod_{\ell=0}^{m_2'-m_2}\hat{Z}_{(m_1,m_2+\ell+\frac{1}{2})}^{a^{m_2'-m_2-\ell}}.\label{e2}
\end{align}
In these expressions, we assumed $0\leq  m_1\leq m_1'\leq L_1-1$ and $0\leq m_2\leq m_2'\leq L_2-1$. 

The state
\begin{align}
\hat{X}_{(m_1-\frac{1}{2},m_2+\frac{1}{2}),(m_1'+\frac{1}{2},m_2+\frac{1}{2})}^{(1)}|\Phi_0\rangle
\end{align}
contains a pair of plaquettes operators with eigenvalues not equal to $1$ (see Fig.~\ref{fig1} (b) for the illustration), which are called magnetic excitations. In this state, the eigenvalues of $\hat{B}_{(m_1-\frac{1}{2},m_2+\frac{1}{2})}$ and $\hat{B}_{(m_1'+\frac{1}{2},m_2+\frac{1}{2})}$ are $\omega$ and $\omega^{-a^{m_1'-m_1+1}}$, respectively. The eigenvalues of other plaquette operators remain $+1$.  In the derivation of these relations, we used the general property of exponents $(z^{m})^n=z^{mn}$ for $z\in\mathbb{C}$ and $m,n\in\mathbb{Z}$. Similarly, the state
\begin{align}
\hat{Z}_{(m_1,m_2),(m_1'+1,m_2)}^{(1)}|\Phi_0\rangle
\end{align}
has the eigenvalues $\omega^{a^{m_1'-m_1+1}}$ and $\omega^{-1}$ for the vertex operators $\hat{A}_{(m_1,m_2)}$ and $\hat{A}_{(m_1'+1,m_2)}$, respectively. 
String operators along $x_2$ direction also create pairs of electric or magnetic excitations at their ends [see Fig.~\ref{fig1} (c)].
  
A single plaquette operator or a vertex operator with eigenvalue $\omega^q$ ($q=1,2,\cdots,N-1$) costs an energy
\begin{align}
&\Delta_q\coloneqq1-\frac{1}{2}(\omega^q+\omega^{-q})=1-\cos\Big(\frac{2\pi q}{N}\Big),\label{singleDelta}\\
&2\geq\Delta_q\geq\Delta_1=1-\cos\Big(\frac{2\pi}{N}\Big).
\end{align}
The excitation energy of a pair $\Delta_{\text{pair}}$ can thus be bounded by 
\begin{align}
&4\geq\Delta_{\text{pair}}\geq 2\Delta_1.\label{pairDelta}
\end{align}

These electric and magnetic excitations can be further divided into equivalence classes up to local excitations (i.e. excitations that can be created locally), which are called the anyon types. They will be discussed in Sec. \ref{sec:anyons}.

\section{Basic facts from number theory}
\label{sec:math}
In this section, we review basic mathematical facts in number theory to setup notations for the following sections.

\subsection{Multiplicative order and primitive root}
\label{mo}
Given a positive integer $n$ and a positive integer $a$ coprime to $n$, the multiplicative order of $a$ modulo $n$ is defined as the smallest positive integer $\ell$ such that 
\begin{equation}
a^\ell=1\mod n,\label{alN}
\end{equation}
which we denote by $M_n(a)$ in this work.
For example, $M_n(a)=1$ if and only if $a=1$ (mod $n$). Also, for $n\geq 3$, $M_n(a)=2$ if $a=-1$ (mod $n$).
Conversely, the relation in Eq.~\eqref{alN} implies that $n$ and $a$ are coprime. In the following applications, the integer $n$ is chosen to be $N$ itself or a divisor of $N$ that is coprime to $a$. 

The multiplicative order is related to Euler's totient function $\varphi(n)$, which is defined as the number of positive integers smaller than $n$ that are relatively prime to $n$. By definition, $1\leq \varphi(n)\leq n-1$. If and only if $n$ is prime, $\varphi(n)=n-1$. 

Euler's theorem~\cite{stein2008elementary}, $a^{\varphi(n)}=1$ mod $n$, implies that 
$M_n(a)$ is a divisor of $\varphi(n)$. Thus
\begin{equation}
1\leq M_n(a)\leq \varphi(n).
\end{equation}
Integers $a$ that saturate the upper bound, i.e., $M_n(a)=\varphi(n)$, are called the primitive roots modulo $n$. The primitive roots exist if and only if $n$ is either $2$, $4$, $p^k$, or $2p^k$, where $p$ is an odd prime number and $k$ is a positive integer.  It follows that, when $n$ is a prime number, there exists an integer $a$ such that 
\begin{align}
M_n(a)=n-1.
\end{align}

Finally, suppose that  $n'$ is also a positive integer coprime to $a$. In this case, $M_{nn'}(a)$ is a multiple of both $M_n(a)$ and $M_{n'}(a)$, because $a^{M_{nn'}(a)}=1$ (mod $nn'$) also implies $a^{M_{nn'}(a)}=1$  (mod $n$) and $a^{M_{nn'}(a)}=1$  (mod $n'$). In particular, 
\begin{align}
M_{n''}(a)\geq M_{n}(a)\label{mnppa}
\end{align}
when $n''$ is a multiple of $n$.

These mathematical facts underlie our results quoted in Eqs.~\eqref{most2} and \eqref{most1}.

\subsection{Prime factorization and divisors of $N$}
Suppose that the integer $N$ ($N\geq 2$) can be prime factorized into
\begin{align}
N=\prod_{j=1}^np_j^{r_j}=p_1^{r_1}p_2^{r_2}\cdots p_n^{r_n},\label{Npf}
\end{align}
where $p_j$'s ($j=1,2,\cdots,n$) are prime numbers and $r_j$'s are positive integers. The radical of $N$ is defined as the product of all distinct prime factors of $N$:
\begin{align}
\mathrm{rad}(N)\coloneqq\prod_{j=1}^np_j=p_1p_2\cdots p_n.
\end{align}
We denote  the set of all (positive) divisors of $N$ by $D_N$:
\begin{align}
D_N=\Big\{\prod_{j=1}^np_j^{r_j'}\Big|0\leq r_j'\leq r_j\Big\},
\end{align}
which includes $1$, $N$, and $\mathrm{rad}(N)$, for example.

Without loss of generality, let us arrange prime factors $p_j$'s of $N$ in Eq.~\eqref{Npf} in such a way that
\begin{align}
\begin{cases}
a/p_j\notin\mathbb{Z}&(j=1,2,\cdots,m),\\
a/p_j\in\mathbb{Z}&(j=m+1,\cdots,n).
\end{cases}
\end{align}
Then the largest divisor of $N$ that is coprime to $a$ is given by
\begin{align}
N_a\coloneqq \prod_{j=1}^{m}p_j^{r_j}=p_1^{r_1}p_2^{r_2}\cdots p_{m}^{r_{m}}\leq N.\label{Na}
\end{align}
By definition, we have
\begin{align}
N_a=N\quad&\Leftrightarrow\quad\mathrm{gcd}(N,a)=1\text{ (i.e., $m=n$)},\\
N_a=1\quad&\Leftrightarrow\quad a/\mathrm{rad}(N)\in\mathbb{Z}\text{ (i.e., $m=0$)}.
\end{align}
Here, $\mathrm{gcd}(p,q,r,\cdots)$ for integers $p, q, r, \cdots$ represents their greatest common divisor. By definition, $p/\mathrm{gcd}(p,q)$ and $q/\mathrm{gcd}(p,q)$ are positive integers. 
Since $N_a$ is a multiple of any $d\in D_N$ that is coprime to $a$, $M_{N_a}(a)$ is a multiple of $M_d(a)$.

\section{Classification of phases in the $(N,a)$ model}
\label{classification}

Our model describes two distinct types of phases with or without topological order depending on whether $a$ ($1\leq a\leq N$) is a multiple of $\mathrm{rad}(N)$ or not. Here we provide a brief summary of the main features of the two phases.

\begin{description}
\item[Case 1] When $a$ is not a multiple of $\mathrm{rad}(N)$, our model exhibits topological degeneracy for some sequences of $L_1$ and $L_2$.
{The ground state degeneracy on the torus is given by
\begin{align}
N_{\mathrm{deg}}&=[\mathrm{gcd}(a^{L_1}-1,a^{L_2}-1,N_a)]^2\label{ndeggeneral}
\end{align}
For example, $N_{\mathrm{deg}}=N_a^2$ when both $L_1$ and $L_2$ are multiples of $M_{N_a}(a)$ and $N_{\mathrm{deg}}=1$ when $L_1$ and $L_2$ are not simultaneously multiples of $M_d(a)$ for any $d\in D_N$ coprime to $a$, except for $d=1$.}  Correspondingly, there are $N_a^2$ species of anyons. This class thus falls into topologically ordered phases. It contains the important class of $a$ being coprime to $N$. Examples include the cases of $a=1$ and $a=N-1$ previously discussed in the literature. The size dependence of $N_{\mathrm{deg}}$ can be understood from the translation symmetry action on the anyon excitations, which will be discussed in Sec. \ref{sec:defect}

\item[Case 2] When $a$ is a multiple of $\mathrm{rad}(N)$, the ground state is unique regardless of the choice of $L_1$ and $L_2$. The model thus realizes a trivial phase with regard to topological orders, but it still might be a nontrivial symmetry protected topological phase. As simplest examples, we discuss the cases of $N=a$ and $N=a^2$.
\end{description}

We study these two cases separately in Secs.~\ref{sec:case1}, \ref{sec:topology} and in Sec.~\ref{sec:case2}.

\section{Ground state degeneracy in topologically ordered phases}
\label{sec:case1}
In this section, we show that, when $a$ is not a multiple of $\mathrm{rad}(N)$, the order of ground state degeneracy $N_{\mathrm{deg}}$ is greater than one for some sequences of $L_1$ and $L_2$.

\subsection{The case of $\mathrm{gcd}(N,a)=1$}
\label{MNa}
We start with the simplest case where $a$ is coprime to $N$.

\subsubsection{When both $L_1$ and $L_2$ are multiples of $M_N(a)$}
\label{MNa1}
Suppose that both $L_1$ and $L_2$ are multiples of $M_N(a)$ so that $a^{L_1}=a^{L_2}=1$ mod $N$. In this case the ground state degeneracy and the low-energy excitations are basically the straightforward extension of the original toric code.
For example, when $a=1$, $M_N(a)=1$ and the assumption automatically holds for any $L_1$ and $L_2$.
 In contrast, when $N\geq3$ and $a=N-1$, $M_N(a)=2$ and both $L_1$ and $L_2$ need to be even.

When both $L_1$ and $L_2$ are multiples of $M_N(a)$, there are two sets of global constraints among the stabilizers $\hat{A}_v$'s and $\hat{B}_p$'s:
\begin{align}
&\prod_{m_1=0}^{L_1-1}\prod_{m_2=0}^{L_2-1}\hat{A}_{(m_1,m_2)}^{a^{m_1+m_2}}\notag\\
&=\prod_{m_1=0}^{L_1-1}\hat{X}_{(m_1,-\frac{1}{2})}^{-a^{m_1}(a^{L_2}-1)}\prod_{m_2=0}^{L_2-1}\hat{X}_{(-\frac{1}{2},m_2)}^{-a^{m_2}(a^{L_1}-1)}=1\label{case0constraint1}
\end{align}
and
\begin{align}
&\prod_{m_1=0}^{L_1-1}\prod_{m_2=0}^{L_2-1}\hat{B}_{(m_1+\frac{1}{2},m_2+\frac{1}{2})}^{a^{(L_1-1-m_1)+(L_2-1-m_2)}}\notag\\
&=\prod_{m_1=0}^{L_1-1}\hat{Z}_{(m_1+\frac{1}{2},0)}^{a^{(L_1-1-m_1)}(a^{L_2}-1)}\prod_{m_2=0}^{L_2-1}\hat{Z}_{(0,m_2+\frac{1}{2})}^{-a^{(L_2-1-m_2)}(a^{L_1}-1)}=1,\label{case0constraint2}
\end{align}
implying that $N_C$ in Eq.~\eqref{Nc} is $N$. In the derivation, we used definitions in Eqs.~\eqref{Av} and~\eqref{Bp} and the periodic boundary condition such as $\hat{X}_{(L_1-\frac{1}{2},m_2)}=\hat{X}_{(-\frac{1}{2},m_2)}$ and $\hat{X}_{(m_1,L_2-\frac{1}{2})}=\hat{X}_{(m_1,-\frac{1}{2})}$.
These constraints imply that not all vertex operators and plaquettes operators are independent. 
For example, the eigenvalues of $\hat{A}_{v_0}$ [$v_0\coloneqq (0,0)$] and $\hat{B}_{p_0}$ [$p_0\coloneqq(L_1-\frac{1}{2},L_2-\frac{1}{2})$] are automatically fixed once the eigenvalues of other $\hat{A}_v$'s and $\hat{B}_p$'s are chosen.  

\begin{figure*}
\includegraphics[width=0.7\textwidth]{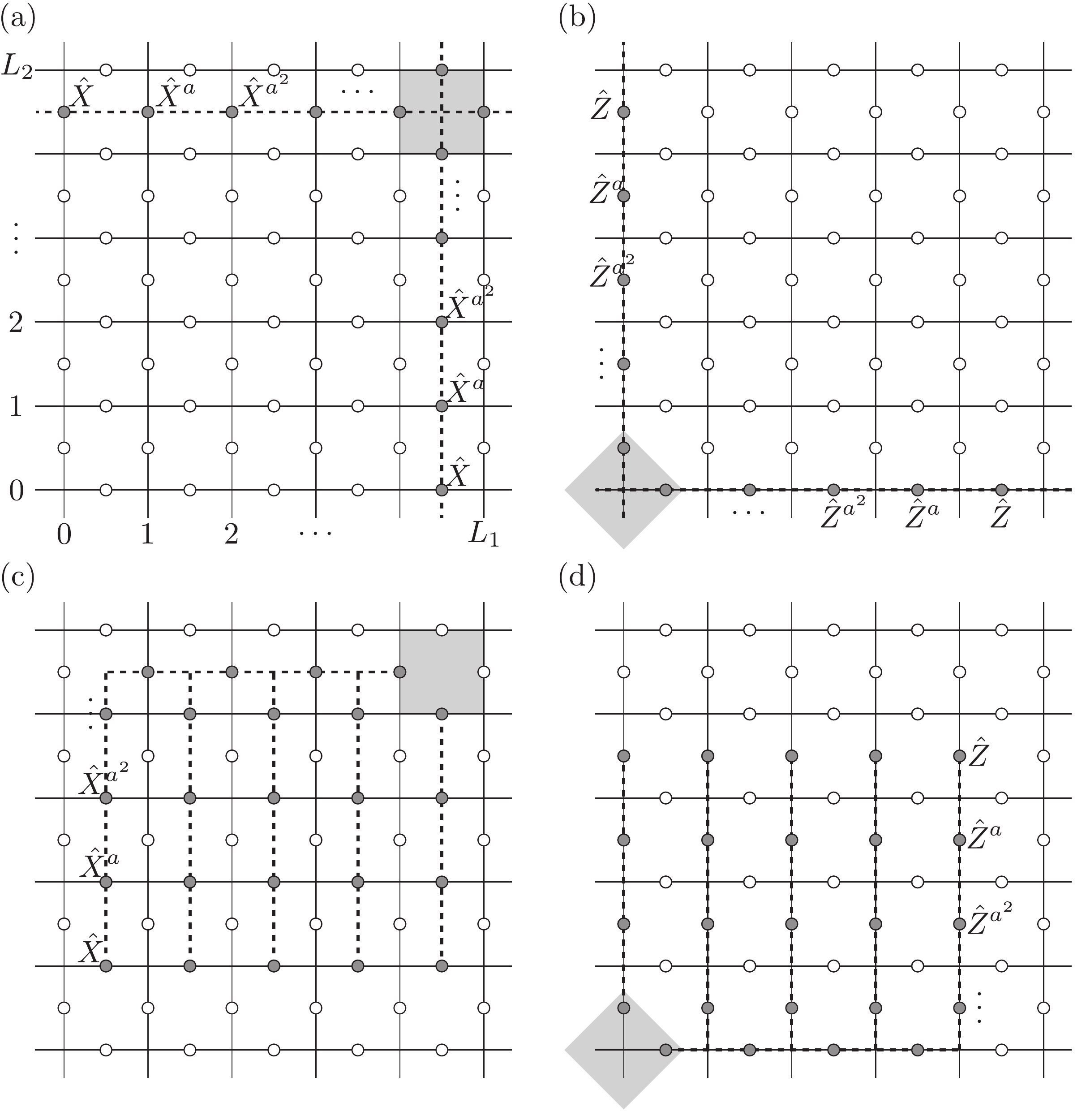}
\caption{\label{fig12} 
Illustration of 
(a) closed string operators $\hat{X}^{(1)}$ and $\hat{X}^{(2)}$, 
(b) closed string operators $\hat{Z}^{(1)}$ and $\hat{Z}^{(2)}$, 
(c) open string operators that control the eigenvalues of plaquette operators $\hat{B}_{p}$ ($p\neq p_0$),
(d) open string operators that control the eigenvalues of vertex operators $\hat{A}_{v}$ ($v\neq v_0$).
}
\end{figure*}

Correspondingly, there are four independent closed string operators, illustrated in Fig.~\ref{fig12} (a) and (b), that commute with every term in the Hamiltonian:
\begin{align}
\hat{X}^{(1)}\coloneqq\hat{X}_{(-\frac{1}{2},L_2-\frac{1}{2}),(L_1-\frac{1}{2},L_2-\frac{1}{2})}^{(1)}&=\prod_{\ell=0}^{L_1-1}\hat{X}_{(\ell,L_2-\frac{1}{2})}^{a^\ell},\label{m12}\\
\hat{X}^{(2)}\coloneqq\hat{X}_{(L_1-\frac{1}{2},-\frac{1}{2}),(L_1-\frac{1}{2},L_2-\frac{1}{2})}^{(2)}&=\prod_{\ell=0}^{L_2-1}\hat{X}_{(L_1-\frac{1}{2},\ell)}^{a^\ell}\label{m22}
\end{align}
and 
\begin{align}
\hat{Z}^{(1)}\coloneqq\hat{Z}_{(0,0),(L_1,0)}^{(1)}&=\prod_{\ell=0}^{L_1-1}\hat{Z}_{(\ell+\frac{1}{2},0)}^{a^{L_1-1-\ell}},\label{e12}\\
\hat{Z}^{(2)}\coloneqq\hat{Z}_{(0,0),(0,L_2)}^{(2)}&=\prod_{\ell=0}^{L_2-1}\hat{Z}_{(0,\ell+\frac{1}{2})}^{a^{L_2-1-\ell}},\label{e22}
\end{align}
where $\hat{X}_{p,p'}^{(i)}$ and $\hat{Z}_{v,v'}^{(i)}$ were defined in Eqs.~\eqref{m1}--\eqref{e2}.
These operators satisfy [see Eq.~\eqref{ZXwXZ}]
\begin{align}
&\hat{Z}^{(1)}\hat{X}^{(2)}=\omega\,\hat{X}^{(2)}\hat{Z}^{(1)},\\
&\hat{Z}^{(2)}\hat{X}^{(1)}=\omega\,\hat{X}^{(1)}\hat{Z}^{(2)},
\end{align}
and
\begin{align}
&[\hat{Z}^{(1)},\hat{Z}^{(2)}]=[\hat{X}^{(1)},\hat{X}^{(2)}]=0,\\
&[\hat{Z}^{(1)},\hat{X}^{(1)}]=[\hat{X}^{(2)},\hat{Z}^{(2)}]=0.
\end{align}
Hence, as the set of independent stabilizers commuting with $\hat{H}$, one can choose the following set of operators:
\begin{itemize}
\item The vertex operators $\hat{A}_v$ ($v\in\mathcal{V}$, $v\neq v_0$) and the plaquette operators $\hat{B}_p$ ($p\in\mathcal{P}$, $p\neq p_0$). There are in total $N^{2(L_1L_2-1)}$ different combinations of eigenvalues.\\
\item Closed string operators $\hat{Z}^{(i)}$ ($i=1,2$). There are $N^2$ different combinations of eigenvalues.
\end{itemize}
Starting from the ground state $|\Phi_0\rangle$ in Eq.~\eqref{gs}, which has the eigenvalue $+1$ for all of these $2L_1L_2$ operators, one can generate all $N^{2L_1L_2}$ states in the Hilbert space by using the open string operators illustrated in Fig.~\ref{fig12} (c) and (d) and the closed loop operators $\hat{X}^{(i)}$ ($i=1,2$). They can be distinguished by $N^{2(L_1L_2-1)}\times N^2=N^{2L_1L_2}$ distinct combinations of eigenvalues of these stabilizers.
In particular, all degenerate ground states can be written as $[\hat{X}^{(1)}]^{j_1}[\hat{X}^{(2)}]^{j_2}|\Phi_0\rangle$ ($j_1,j_2=0,1,\cdots,N-1$), which has the eigenvalue $\omega^{j_i}$ of $\hat{Z}^{(i)}$.
Hence, the order of topological degeneracy is
\begin{align}
N_{\mathrm{deg}}=N^2.
\end{align}

The closed loop operators in Eqs.~\eqref{m12} and \eqref{m22} create a pair of magnetic excitations at $x_i=\pm\frac{1}{2}$, dragging the one at $x_i=\frac{1}{2}$ all the way to $x_i=L_i-\frac{1}{2}=-\frac{1}{2}$, and annihilating them in pair. The pair annihilation requires that magnetic excitations with eigenvalues $\omega$ and $\omega^{-1}$ meet. This is possible only when $L_i$ is a multiple of $M_N(a)$.

\subsubsection{When $L_1$ or $L_2$ is not a multiple of $M_N(a)$}
\label{sec:unique}
Next, we consider the case where $L_1$ or $L_2$ is not a multiple of $M_N(a)$.  
Without loss of generality, we assume that $L_1$ is not a multiple of $M_N(a)$. 

Let us introduce a product of string operators associated with the plaquette $p=(m_1+\tfrac{1}{2},m_2+\tfrac{1}{2})\in\mathcal{P}$:
\begin{align}
\hat{X}_p^{(1)}&\coloneqq
\Big[\hat{X}_{(-\frac{1}{2},m_2+\frac{1}{2}),p}^{(1)}\Big]^{a^{L_1-1-m_1}}\hat{X}_{p,(L_1-\frac{1}{2},m_2+\frac{1}{2})}^{(1)}.\label{Xp1}
\end{align}
The first factor creates magnetic excitations with eigenvalues $\omega^{a^{L_1-1-m_1}}$ and $\omega^{-a^{L_1}}$ at the plaquettes $(-\frac{1}{2},m_2+\frac{1}{2})$ and $p$, respectively.
The second factor creates magnetic excitations with eigenvalues $\omega$ and $\omega^{-a^{L_1-1-m_1}}$ at the plaquettes $p$ and $(L_1-\frac{1}{2},m_2+\frac{1}{2})$, respectively. Combining these two effects, the operator $\hat{X}_p^{(1)}$ create a \emph{single} magnetic excitation with eigenvalue $\omega^{1-a^{L_1}}$ at the plaquette $p$.  In fact, $\hat{X}_p^{(1)}$ satisfies
\begin{align}
&\hat{A}_v\hat{X}_p^{(1)}=\hat{X}_p^{(1)}\hat{A}_v,\\
&\hat{B}_{p'}\hat{X}_p^{(1)}=\omega^{(1-a^{L_1})\delta_{p,p'}}\hat{X}_p^{(1)}\hat{B}_{p'}
\end{align}
for any $v\in\mathcal{V}$ and $p'\in\mathcal{P}$. Since $L_1$ is not a multiple of $M_N(a)$, $\omega^{1-a^{L_1}}\neq 1$.

Similarly, the following operator can be introduced for each vertex $v=(m_1,m_2)\in\mathcal{V}$:
\begin{align}
\hat{Z}_v^{(1)}\coloneqq
\hat{Z}_{(0,m_2),v}^{(1)}
\Big[\hat{Z}_{v,(L_1,m_2)}^{(1)}\Big]^{a^{m_1}}\label{Zv1}
\end{align}
which satisfies
\begin{align}
&\hat{A}_{v'}\hat{Z}_v^{(1)}=\omega^{(a^{L_1}-1)\delta_{v,v'}}\hat{Z}_v^{(1)}\hat{A}_{v'},\\
&\hat{B}_p\hat{Z}_v^{(1)}=\hat{Z}_v^{(1)}\hat{B}_p
\end{align}
for any $v'\in\mathcal{V}$ and $p\in\mathcal{P}$.  Hence, $\hat{Z}_v^{(1)}$ creates  a \emph{single} electric excitation with eigenvalue $\omega^{a^{L_1}-1}\neq1$ at the vertex $v$. 

To proceed, let us assume further that $a^{L_1}-1$ is coprime to $N$. In this case
\begin{align}
\ell(a^{L_1}-1)\mod N\quad (\ell=1,2,\cdots,N-1)
\end{align}
are all different and are not equal to $0$ mod $N$. Therefore, the eigenvalue of the plaquette operator $\hat{B}_p$ (the vertex operator $\hat{A}_v$) can be freely controlled by $[\hat{X}_p^{(1)}]^\ell$ ($[\hat{Z}_v^{(1)}]^\ell$) without affecting others, implying the absence of global constraints involving $\hat{B}_p$ or $\hat{A}_v$, such as the ones of the form in Eq.~\eqref{constraint} (i.e, $N_C=1$).

Moreover, operators $\hat{X}_v^{(1)}$ ($v\in\mathcal{V}$) and $\hat{Z}_p^{(1)}$ ($p\in\mathcal{P}$) all commute with each other. Hence, starting from the ground state $|\Phi_0\rangle$ satisfying Eqs.~\eqref{AvPhi} and~\eqref{BpPhi}, one can generate all $N^{2L_1L_2}$ states in the Hilbert space by successively applying $\hat{Z}_v^{(1)}$'s and $\hat{X}_p^{(1)}$'s. In particular, there is no state other than $|\Phi_0\rangle$ that has eigenvalue $+1$ for all vertex operators and plaquette operators. This proves the uniqueness of the ground state 
\begin{align}
N_{\mathrm{deg}}=1,
\end{align}
given that $a^{L_1}-1$ is coprime to $N$. This condition is satisfied, for example, (i) when $N$ is prime and $L_1$ is not a multiple of $M_N(a)$ (in this case $a^{L_1}-1\neq0$ mod $N$) and (ii) when $N$ is odd, $a=N-1$, and $L_1$ is not a multiple of $M_N(a)=2$ (in this case $a^{L_1}-1=-2$ mod $N$).  This completes the proof of Eqs.~\eqref{most2} and~\eqref{most1}.

The gap to the first excited states are given by $\Delta_q$ in Eq.~\eqref{singleDelta}, although these states are created by nonlocal operators $\hat{X}_p^{(1)}$ and $\hat{Z}_v^{(1)}$. Local excitations are still given by pairs of magnetic excitations and electric excitations, for which the exaction gap is bounded by Eq.~\eqref{pairDelta}.

\subsection{General case}
\label{sec:general}
Next we discuss the most general case where  $a/\mathrm{rad}(N)\notin\mathbb{Z}$ but $a$ is not necessarily coprime to $N$.
In this case, we will see that $N_a$ in Eq.~\eqref{Na} plays the role of $N$ in the above discussion.

Let us list up all constraints among $\hat{A}_v$'s and $\hat{B}_p$'s of the form of Eqs.~\eqref{constraint} and ~\eqref{constraint2}. We have
\begin{align}
&\prod_{m_1=0}^{L_1-1}\prod_{m_2=0}^{L_2-1}\hat{A}_{(m_1,m_2)}^{na^{m_1+m_2}}\notag\\
&=\prod_{m_1=0}^{L_1-1}\hat{X}_{(m_1,-\frac{1}{2})}^{-a^{m_1}(a^{L_2}-1)n}\prod_{m_2=0}^{L_2-1}\hat{X}_{(-\frac{1}{2},m_2)}^{-a^{m_2}(a^{L_1}-1)n}\label{case1constraint3}
\end{align}
and
\begin{align}
&\prod_{m_1=0}^{L_1-1}\prod_{m_2=0}^{L_2-1}\hat{B}_{(m_1+\frac{1}{2},m_2+\frac{1}{2})}^{na^{(L_1-1-m_1)+(L_2-1-m_2)}}\notag\\
&=\prod_{m_1=0}^{L_1-1}\hat{Z}_{(m_1+\frac{1}{2},0)}^{a^{(L_1-1-m_1)}(a^{L_2}-1)n}\prod_{m_2=0}^{L_2-1}\hat{Z}_{(0,m_2+\frac{1}{2})}^{-a^{(L_2-1-m_2)}(a^{L_1}-1)n}.\label{case1constraint4}
\end{align}
Here, $n\in D_N$ is a parameter specified shortly. In order to set the products in Eqs.~\eqref{case1constraint3} and \eqref{case1constraint4} to be $1$, we need
\begin{align}
(a^{L_1}-1)n=(a^{L_2}-1)n=0\mod N.\label{aLi2}
\end{align}
To solve this equation, let us define $d_{i,a}\in D_N$ $(i=1,2)$ by
\begin{align}
d_{i,a}\coloneqq\mathrm{gcd}(a^{L_i}-1,N_a).
\end{align}
This is the largest divisor of $N$ such that (i) $d$ is coprime to $a$ and (ii) $L_i$ is a multiple of $M_{d}(a)$.
For example, $d_{i,a}=N$ when $a=1$, and $d_{i,a}=1$ when $a^{L_i}-1$ is coprime to $N$.  The smallest positive integer $n\in D_N$ satisfying Eq.~\eqref{aLi2} is given by 
\begin{align}
&n=n_a\coloneqq \frac{N}{d_a},
\end{align}
where
\begin{align}
&d_a\coloneqq\mathrm{gcd}(d_{1,a},d_{2,a})=\mathrm{gcd}(a^{L_1}-1,a^{L_2}-1,N_a).\label{defda}
\end{align}

After all, we find the following set of global constraints
\begin{align}
&\left(\prod_{m_1=0}^{L_1-1}\prod_{m_2=0}^{L_2-1}\hat{A}_{(m_1,m_2)}^{ n_aa^{m_1+m_2}}\right)^{n'}=1\label{case1constraint1},\\
&\left(\prod_{m_1=0}^{L_1-1}\prod_{m_2=0}^{L_2-1}\hat{B}_{(m_1+\frac{1}{2},m_2+\frac{1}{2})}^{n_aa^{(L_1-1-m_1)+(L_2-1-m_2)}}\right)^{n'}=1\label{case1constraint2},
\end{align}
where $n'=0,1,2,\cdots,d_a-1$, suggesting that $N_C$ in Eq.~\eqref{Nc} is given by $d_a$.  These constraints imply that not all vertex operators and plaquettes operators are independent. 
For example, the eigenvalues of $\hat{A}_{v_0}^{n_a}$ and $\hat{B}_{p_0}^{n_a}$ can be automatically fixed 
once the eigenvalues of other $\hat{A}_v$'s and $\hat{B}_p$'s are chosen.  Then, as the set of independent stabilizers commuting with $\hat{H}$, one can choose the following set of operators:
\begin{itemize}
\item The vertex operators $\hat{A}_v$ ($v\in\mathcal{V}$, $v\neq v_0$) and the plaquette operators $\hat{B}_p$ ($p\in\mathcal{P}$, $p\neq p_0$). There are in total $N^{2(L_1L_2-1)}$ different combinations of eigenvalues.\\
\item The residual free parts of $\hat{A}_{v_0}$ and $\hat{B}_{p_0}$. The eigenvalues of these operators can be written as $\omega^{x+d_a\ell}$ ($\ell=0,1,\cdots,n_a-1$), where the value of $x$ ($x=0,1,\cdots,d_a$) is automatically determined by the constraints in Eqs.~\eqref{case1constraint1} and \eqref{case1constraint2}. Hence, there are effectively $n_a^2$ different combinations of eigenvalues.\\
\item Loop operators $[\hat{X}^{(1)}]^{n_{1,a}}$ and $[\hat{Z}^{(1)}]^{n_{1,a}}$ (or $[\hat{X}^{(2)}]^{n_{2,a}}$ and $[\hat{Z}^{(2)}]^{n_{2,a}}$), where $n_{i,a}\coloneqq N/d_{i,a}$ ($i=1,2$). Their eigenvalues are $d_{i,a}$-fold: $\omega^{n_{i,a}j}$ ($j=0,1,\cdots,d_{i,a}-1$), which include $\omega^{n_aj'}$  $(j'=0,1,\cdots,d_{a}-1)$ as a subset. As detailed below, only $d_a^2$ different eigenvalues of these operators can be manipulated without affecting the eigenvalues of other stabilizers. 
\end{itemize}
Hence, starting from the ground state $|\Phi_0\rangle$ in Eq.~\eqref{gs}, one can generate all $N^{2L_1L_2}$ states in the Hilbert space, which can be distinguished by $N^{2(L_1L_2-1)}\times n_a^2\times d_a^2=N^{2L_1L_2}$ distinct combinations of eigenvalues of these stabilizers. This implies that the order of the ground state degeneracy is given by Eq.~\eqref{ndeggeneral}. 

It remains to show that the eigenvalues of stabilizers can be manipulated as stated above. Clearly, open string operators illustrated in Fig.~\ref{fig12} (c) and (d) can be used to control the eigenvalues of $\hat{A}_v$ ($v\in\mathcal{V}$, $v\neq v_0$) and $\hat{B}_p$ ($p\in\mathcal{P}$, $p\neq p_0$). The remaining operators satisfy the following algebra:
\begin{align}
&\hat{Z}^{(i)}\hat{A}_{v_0}=\omega^{\alpha_i}\hat{A}_{v_0}\hat{Z}^{(i)},\\
&\hat{X}^{(i)}\hat{B}_{p_0}=\omega^{-\alpha_i}\hat{B}_{p_0}\hat{X}^{(i)},\\
&\hat{Z}^{(1)}\hat{X}^{(2)}=\omega\,\hat{X}^{(2)}\hat{Z}^{(1)},\\
&\hat{Z}^{(2)}\hat{X}^{(1)}=\omega\,\hat{X}^{(1)}\hat{Z}^{(2)},
\end{align}
where $\alpha_i$ ($0\leq \alpha_i\leq N-1$) is defined by 
\begin{align}
\alpha_i= a^{L_i}-1\mod N,
\end{align}
which is coprime to $a$ and a multiple of $d_{i,a}=\mathrm{gcd}(\alpha_i,N_a)$.  All of these operators commute with $\hat{A}_v$ ($v\in\mathcal{V}$, $v\neq v_0$) and $\hat{B}_p$ ($p\in\mathcal{P}$, $p\neq p_0$) and thus do not change their eigenvalues.

\subsubsection{Case 1: $\alpha_1=\alpha_2=0$}
When $\alpha_1=\alpha_2=0$, $N$ is coprime to $a$ and both $L_1$ and $L_2$ are multiples of $M_N(a)$.  
This case was covered in Sec.~\ref{MNa1}.

\subsubsection{Case 2: Either $\alpha_1= 0$ or $\alpha_2=0$}
Next we discuss the case when either $\alpha_1= 0$ or $\alpha_2=0$. Without loss of the generality, here we assume $\alpha_1\neq0$ and $\alpha_2=0$. 
In this case, $N$ is again coprime to $a$, and we have $d_{1,a}=d_a=\mathrm{gcd}(\alpha_1,N)$ and $d_{2,a}=N$.

Since $\alpha_1/d_{1,a}$ is coprime to $n_{1,a}= N/d_{1,a}$, there exists an integer $\ell_1$ ($1\leq \ell_1\leq n_{1,a}-1$) such that
\begin{align}
\ell_1\frac{\alpha_1}{d_{1,a}}=1\mod n_{1,a}.
\end{align}
Then, we can control the eigenvalues of $\hat{A}_{v_0}$ and $\hat{B}_{p_0}$ by $[\hat{X}^{(1)}]^{\ell_1}$ and $[\hat{Z}^{(1)}]^{\ell_1}$:
\begin{align}
&[\hat{X}^{(1)}]^{\ell_1}\hat{A}_{v_0}=\hat{A}_{v_0}[\hat{X}^{(1)}]^{\ell_1},\\
&[\hat{Z}^{(1)}]^{\ell_1}\hat{A}_{v_0}=\omega^{d_a}\hat{A}_{v_0}[\hat{Z}^{(1)}]^{\ell_1},\\
&[\hat{X}^{(1)}]^{\ell_1}\hat{B}_{p_0}=\omega^{-d_a}\hat{B}_{p_0}[\hat{X}^{(1)}]^{\ell_1},\\
&[\hat{Z}^{(1)}]^{\ell_1}\hat{B}_{p_0}=\hat{B}_{p_0}[\hat{Z}^{(1)}]^{\ell_1}
\end{align}
without affecting the eigenvalues of $[\hat{X}^{(1)}]^{n_{1,a}}$ and $[\hat{Z}^{(1)}]^{n_{1,a}}$. We can also control the eigenvalues of $[\hat{X}^{(1)}]^{n_{1,a}}$ and $[\hat{Z}^{(1)}]^{n_{1,a}}$ by $\hat{X}^{(2)}$ and $\hat{Z}^{(2)}$:
\begin{align}
&\hat{X}^{(2)}[\hat{X}^{(1)}]^{n_{1,a}}=[\hat{X}^{(1)}]^{n_{1,a}}\hat{X}^{(2)},\\
&\hat{Z}^{(2)}[\hat{X}^{(1)}]^{n_{1,a}}=\omega^{n_{1,a}}\,[\hat{X}^{(1)}]^{n_{1,a}}\hat{Z}^{(2)},\\
&\hat{X}^{(2)}[\hat{Z}^{(1)}]^{n_{1,a}}=\omega^{-n_{1,a}}\,[\hat{Z}^{(1)}]^{n_{1,a}}\hat{X}^{(2)},\\
&\hat{Z}^{(2)}[\hat{Z}^{(1)}]^{n_{1,a}}=[\hat{Z}^{(1)}]^{n_{1,a}}\hat{Z}^{(2)}
\end{align}
without affecting the eigenvalues of $\hat{A}_{v_0}$ and $\hat{B}_{p_0}$. Since $n_{1,a}=n_a=N/d_a$, this is what we needed.

\subsubsection{Case 3: $\alpha_1\neq 0$ and $\alpha_2\neq0$}

Finally, we discuss the case when $\alpha_1\neq0$ and $\alpha_2\neq0$. 
We define operators $\hat{X}^{(\ell_1,\ell_2)}\coloneqq [\hat{X}^{(1)}]^{\ell_1}[\hat{X}^{(2)}]^{\ell_2}$ and $\hat{Z}^{(\ell_1,\ell_2)}\coloneqq [\hat{Z}^{(1)}]^{\ell_1}[\hat{Z}^{(2)}]^{\ell_2}$.

Since $a$ and $\alpha_i$ are coprime, $d_a$ in Eq.~\eqref{defda} can also be written as $\mathrm{gcd}(\alpha_1,\alpha_2,N)$.
It follows that $\mathrm{gcd}(\alpha_1,\alpha_2)/d_{a}$ is coprime to $n_a=N/d_a$. Thus there exists an integer $b_0$ such that
\begin{align}
b_0\frac{\mathrm{gcd}(\alpha_1,\alpha_2)}{d_a}=1\mod n_a.
\end{align}
Furthermore, B\'ezout's lemma tells us the existence of integers $b_1$ and $b_2$ such that
\begin{align}
b_1\alpha_1+b_2\alpha_2=\mathrm{gcd}(\alpha_1,\alpha_2).
\end{align}
Therefore, we have
\begin{align}
&\ell_1\frac{\alpha_1}{d_a}+\ell_2\frac{\alpha_2}{d_a}=1\mod n_a
\end{align}
with $\ell_i=b_0b_i$ mod $n_a$ ($0\leq \ell_i\leq n_a-1$). 
The eigenvalues of $\hat{A}_{v_0}$ and $\hat{B}_{p_0}$ can be controlled by $\hat{X}^{(\ell_1,\ell_2)}$ and $\hat{Z}^{(\ell_1,\ell_2)}$:
\begin{align}
&\hat{X}^{(\ell_1,\ell_2)}\hat{A}_{v_0}=\hat{A}_{v_0}\hat{X}^{(\ell_1,\ell_2)},\\
&\hat{Z}^{(\ell_1,\ell_2)}\hat{A}_{v_0}=\omega^{d_a}\hat{A}_{v_0}\hat{Z}^{(\ell_1,\ell_2)},\\
&\hat{X}^{(\ell_1,\ell_2)}\hat{B}_{p_0}=\omega^{-d_a}\hat{B}_{p_0}\hat{X}^{(\ell_1,\ell_2)},\\
&\hat{Z}^{(\ell_1,\ell_2)}\hat{B}_{p_0}=\hat{B}_{p_0}\hat{Z}^{(\ell_1,\ell_2)}.
\end{align}
This process might affect the eigenvalues of the closed loop operators $[\hat{X}^{(i)}]^{n_{i,a}}$ and $[\hat{Z}^{(i)}]^{n_{i,a}}$.

Next, suppose that
\begin{align}
&\ell_1'\frac{\alpha_1}{d_a}+\ell_2'\frac{\alpha_2}{d_a}=0\mod n_a.
\end{align}
In this case, $\hat{X}^{(\ell_1',\ell_2')}$ and $\hat{Z}^{(\ell_1',\ell_2')}$ commute with $\hat{A}_{v_0}$ and $\hat{B}_{p_0}$. 
For example, one can set $\ell_1'=-(\alpha_2+b_2N)/\mathrm{gcd}(\alpha_1+b_1N,\alpha_2+b_2N)$ and $\ell_2'=(\alpha_1+b_1N)/\mathrm{gcd}(\alpha_1+b_1N,\alpha_2+b_2N)$ with $b_1,b_2\in\mathbb{Z}$ being free parameters. 
Choosing $\ell_1'$ and $\ell_2'$ properly, we can realize
\begin{align}
\mathrm{gcd}\left(\frac{n_{1,a}}{n_{a}}\ell_2',d_a\right)=1.\label{conjecture}
\end{align}
See Appendix~\ref{app:proof} for the proof.
Assuming this and using the relations
\begin{align}
&\hat{X}^{(\ell_1',\ell_2')}[\hat{X}^{(1)}]^{n_{1,a}}=[\hat{X}^{(1)}]^{n_{1,a}}\hat{X}^{(\ell_1',\ell_2')},\\
&\hat{X}^{(\ell_1',\ell_2')}[\hat{Z}^{(1)}]^{n_{1,a}}=\omega^{-n_{1,a}\ell_2'}\,[\hat{Z}^{(1)}]^{n_{1,a}}\hat{X}^{(\ell_1',\ell_2')},\\
&\hat{Z}^{(\ell_1',\ell_2')}[\hat{X}^{(1)}]^{n_{1,a}}=\omega^{n_{1,a}\ell_2'}\,[\hat{X}^{(1)}]^{n_{1,a}}\hat{Z}^{(\ell_1',\ell_2')},\\
&\hat{Z}^{(\ell_1',\ell_2')}[\hat{Z}^{(1)}]^{n_{1,a}}=[\hat{Z}^{(1)}]^{n_{1,a}}\hat{Z}^{(\ell_1',\ell_2')},
\end{align}
we can control the eigenvalues of $[\hat{X}^{(1)}]^{n_{1,a}}$ and $[\hat{Z}^{(1)}]^{n_{1,a}}$ by a multiple of $\omega^{n_{a}}$ without affecting the eigenvalues of $\hat{A}_{v_0}$ and $\hat{B}_{p_0}$.
This completes the proof of Eq.~\eqref{ndeggeneral}.

\section{Relation to the standard $\mathbb{Z}_N$ toric code}
\label{sec:relation}

In this section, we clarify the relation of our model in Eq.~\eqref{defH} to the $a=1$ $\mathbb{Z}_N$ toric code with twisted boundary condition. 
This connection for a prime $N$ is implied by the result in Ref.~\onlinecite{doi:10.1063/5.0021068}, but our discussion goes more generally whenever $N$ and $a$ are coprime.

Let us consider a modified Hamiltonian
\begin{align}
\hat{H}'&=-\sum_{v\in\mathcal{V}}\sum_{\ell=0}^{M_N(a)-1}\frac{1}{2}(\hat{A}_v^{a^\ell}+\text{h.c.})-\sum_{p\in\mathcal{P}}\sum_{\ell=0}^{M_N(a)-1}\frac{1}{2}(\hat{B}_p^{a^\ell}+\text{h.c.}),
\end{align}
We still assume the periodic boundary condition. This model is equivalent to $\hat{H}$ in Eq.~\eqref{defH} in the sense that it is written as the sum of the same set of stabilizers $\hat{A}_v$ ($v\in\mathcal{V}$) and $\hat{B}_p$ ($p\in\mathcal{P}$) in Eqs.~\eqref{Av} and \eqref{Bp} with $a_1=a_2=a$. 
The ground states are still given by those who have eigenvalue $+1$ for all $\hat{A}_v$ ($v\in\mathcal{V}$) and $\hat{B}_p$ ($p\in\mathcal{P}$)
and the ground state degeneracy remains unchanged. 


We introduce a local unitary operator $\hat{U}_{\bm{r}}$ ($\bm{r}\in\Lambda$), whose action on the local spin is given by a unitary matrix $U_{i,j}\coloneqq\delta_{j,1+\text{mod}[(i-1)a,N]}$.
This operator satisfies
\begin{align}
&\hat{U}_{\bm{r}}\hat{X}_{\bm{r}}^{a^\ell}\hat{U}_{\bm{r}}^\dagger=\hat{X}_{\bm{r}}^{a^{\ell-1}},\\
&\hat{U}_{\bm{r}}\hat{Z}_{\bm{r}}^{a^\ell}\hat{U}_{\bm{r}}^\dagger=\hat{Z}_{\bm{r}}^{a^{\ell+1}},\\
&\hat{U}_{\bm{r}}^{M_N(a)}=1.
\end{align}
Here and hereafter, $\hat{X}_{\bm{r}}^{\pm a^{-\ell}}$ ($\ell=1,2,\cdots,M_N(a)$) should be understood as $\hat{X}_{\bm{r}}^{\pm a^{M_N(a)-\ell}}$ (recall that $a^{M_N(a)}=1$ mod $N$).  The global operator $\prod_{\bm{r}\in\Lambda}\hat{U}_{\bm{r}}$ is a symmetry of $\hat{H}'$ as it commutes with $\hat{H}'$.  When $\mathrm{gcd}(N,a)\neq1$, such a unitary operator does not exist. 

\begin{figure}
\includegraphics[width=0.5\textwidth]{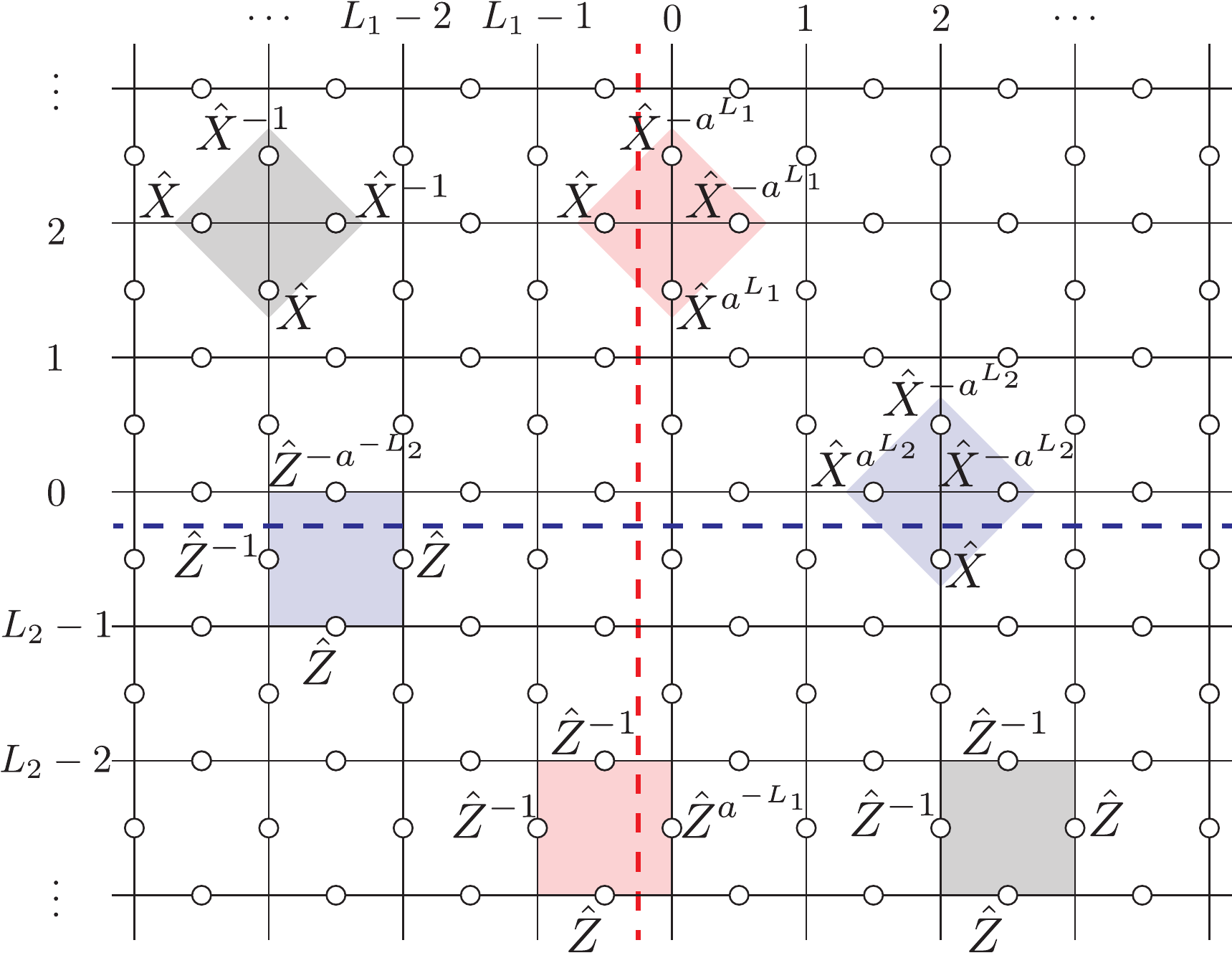}
\caption{\label{figd2}
The standard $\mathbb{Z}_N$ toric code with twisted boundary condition.}
\end{figure}

Now, let us define a twist operator
\begin{align}
\hat{U}\coloneqq \prod_{m_1=0}^{L_1-1}\prod_{m_2=0}^{L_2-1}[\hat{U}_{(m_1+\frac{1}{2},m_2)} \hat{U}_{(m_1,m_2+\frac{1}{2})}]^{m_1+m_2}.
\end{align}
The twist operator convert the stabilizers $\hat{A}_v$ and $\hat{B}_p$ away from the boundary (i.e., $1\leq m_1\leq L_1-2$ and $1\leq m_2\leq L_2-2$) to those for $a=1$:
\begin{align}
&\hat{U}\hat{A}_{(m_1,m_2)}\hat{U}^\dagger=\big(\hat{A}_{(m_1,m_2)}^{(1)}\big)^{a^{1-m_1-m_2}},\\
&\hat{U}\hat{B}_{(m_1+\frac{1}{2},m_2+\frac{1}{2})}\hat{U}^\dagger=\big(\hat{B}_{(m_1+\frac{1}{2},m_2+\frac{1}{2})}^{(1)}\big)^{a^{1+m_1+m_2}}.
\end{align}
Here, $\hat{A}_v^{(1)}$ and $\hat{B}_p^{(1)}$ represent $\hat{A}_v$ and $\hat{B}_p$ in Eqs.~\eqref{Av} and \eqref{Bp} for $a_1=a_2=1$, respectively.
Therefore, except for boundary terms,
\begin{align}
\hat{U}\hat{H}'\hat{U}^\dagger&=-\sum_{v\in\mathcal{V}}\sum_{\ell=0}^{M_N(a)-1}\frac{1}{2}\Big[\big(\hat{A}_v^{(1)}\big)^{a^\ell}+\text{h.c.}\Big]\notag\\
&\quad-\sum_{p\in\mathcal{P}}\sum_{\ell=0}^{M_N(a)-1}\frac{1}{2}\Big[\big(\hat{B}_p^{(1)}\big)^{a^\ell}+\text{h.c.}\Big]
\end{align}
is equivalent to the standard $\mathbb{Z}_N$ toric code ($a=1$). Boundary terms are given by
\begin{align}
&\hat{A}_{(0,m_2)}^{(1)}\coloneqq\hat{X}_{(\frac{1}{2},m_2)}^{-a^{L_1}}\hat{X}_{(0,m_2+\frac{1}{2})}^{-a^{L_1}}\hat{X}_{(L_1-\frac{1}{2},m_2)}\hat{X}_{(0,m_2-\frac{1}{2})}^{a^{L_1}},\\
&\hat{A}_{(m_1,0)}^{(1)}\coloneqq\hat{X}_{(m_1+\frac{1}{2},0)}^{-a^{L_2}}\hat{X}_{(m_1,\frac{1}{2})}^{-a^{L_2}}\hat{X}_{(m_1-\frac{1}{2},0)}^{a^{L_2}}\hat{X}_{(m_1,L_2-\frac{1}{2})},\\
&\hat{A}_{(0,0)}^{(1)}\coloneqq\hat{X}_{(\frac{1}{2},0)}^{-a^{L_1+L_2}}\hat{X}_{(0,\frac{1}{2})}^{-a^{L_1+L_2}}\hat{X}_{(L_1-\frac{1}{2},0)}^{a^{L_2}}\hat{X}_{(0,L_2-\frac{1}{2})}^{a^{L_1}}
\end{align}
and
\begin{align}
&\hat{B}_{(L_1-\frac{1}{2},m_2+\frac{1}{2})}^{(1)}\coloneqq\hat{Z}_{(0,m_2+\frac{1}{2})}^{a^{-L_1}}\hat{Z}_{(L_1-\frac{1}{2},m_2+1)}^{-1}\hat{Z}_{(L_1-1,m_2+\frac{1}{2})}^{-1}\hat{Z}_{(L_1-\frac{1}{2},m_2)},\\
&\hat{B}_{(m_1+\frac{1}{2},L_2-\frac{1}{2})}^{(1)}\coloneqq\hat{Z}_{(m_1+1,L_2-\frac{1}{2})}\hat{Z}_{(m_1+\frac{1}{2},0)}^{-a^{-L_2}}\hat{Z}_{(m_1,L_2-\frac{1}{2})}^{-1}\hat{Z}_{(m_1+\frac{1}{2},L_2-1)},\\
&\hat{B}_{(L_1-\frac{1}{2},L_2-\frac{1}{2})}^{(1)}\coloneqq\hat{Z}_{(0,L_2-\frac{1}{2})}^{a^{-L_1}}\hat{Z}_{(L_1-\frac{1}{2},0)}^{-a^{-L_2}}\hat{Z}_{(L_1-1,L_2-\frac{1}{2})}^{-1}\hat{Z}_{(L_1-\frac{1}{2},L_2-1)}.
\end{align}
See Fig~\ref{figd2} for the illustration. These boundary terms can be understood as a result of twisted boundary condition
\begin{align}
&\hat{X}_{\bm{r}+(L_1,0)}=(\hat{U}^\dagger)^{L_1}\hat{X}_{\bm{r}}\hat{U}^{L_1}=\hat{X}_{\bm{r}}^{a^{L_1}},\\
&\hat{Z}_{\bm{r}+(L_1,0)}=(\hat{U}^\dagger)^{L_1}\hat{Z}_{\bm{r}}\hat{U}^{L_1}=\hat{Z}_{\bm{r}}^{a^{-L_1}},\\
&\hat{X}_{\bm{r}+(0,L_2)}=(\hat{U}^\dagger)^{L_2}\hat{X}_{\bm{r}}\hat{U}^{L_2}=\hat{X}_{\bm{r}}^{a^{L_2}},\\
&\hat{Z}_{\bm{r}+(0,L_2)}=(\hat{U}^\dagger)^{L_2}\hat{Z}_{\bm{r}}\hat{U}^{L_2}=\hat{Z}_{\bm{r}}^{a^{-L_2}}.
\end{align}
This boundary condition modifies the translation symmetries to $\hat{T}_1'\coloneqq\prod_{m_2=0}^{L_2-1}[\hat{U}_{(\frac{1}{2},m_2)}^\dagger \hat{U}_{(0,m_2+\frac{1}{2})}^\dagger]^{L_1}\hat{T}_1$ and $\hat{T}_2'\coloneqq\prod_{m_1=0}^{L_1-1}[\hat{U}_{(m_1,\frac{1}{2})}^\dagger \hat{U}_{(m_1+\frac{1}{2},0)}^\dagger]^{L_2}\hat{T}_2$ and the original translation symmetries in Eq.~\eqref{translation} are broken.

When $N$ and $a$ are not coprime, our model in Eq.~\eqref{defH} cannot be mapped to the standard $\mathbb{Z}_N$ toric code in this way. In fact, as we shall see in the next section, they have different topological orders and cannot be mapped to each other by local unitary transformations.

\section{Topological properties  in topologically ordered phases}
\label{sec:topology}
In Sec.~\ref{sec:case1}, we showed that the order of ground state degeneracy under the periodic boundary condition can be $1$  depending on the system size. Then one might suspect that the system is not in a topologically ordered phase. In this section, we show this is not the case by demonstrating nontrivial topological entanglement entropy and anyonic excitations in the system. In addition, we discuss how the size dependence of the ground state degeneracy can be understood by viewing the lattice system as a continuum torus but with lattice translation symmetry defects.

\subsection{Topological entanglement entropy}
Here we compute the topological entanglement entropy $S_{\mathrm{topo}}$ of the ground state of our model.
We use the Kitaev--Preskill prescription\cite{KitaevPreskill}
\begin{equation}
S_{\mathrm{topo}}=(S_{\mathrm{A}}+S_{\mathrm{B}}+S_{\mathrm{C}})-(S_{\mathrm{AB}}+S_{\mathrm{BC}}+S_{\mathrm{CA}})+S_{\mathrm{ABC}},\label{KP}
\end{equation}
where
\begin{align}
&S_{\mathrm{R}}\coloneqq-\mathrm{tr}[\hat{\rho}_{\mathrm{R}}\log\hat{\rho}_{\mathrm{R}}]
\end{align}
is the von Neumann entropy of the subregion $\mathrm{R}$ of the system and $\hat{\rho}_{\mathrm{R}}\coloneqq\mathrm{tr}_{\bar{\mathrm{R}}}|\Phi_0\rangle\langle\Phi_0|$ ($\mathrm{tr}_{\bar{\mathrm{R}}}$ represents the partial trace over the complement of the region $\mathrm{R}$) is the reduced density matrix of the ground state $|\Phi_0\rangle$. The von Neumann entropy shows the area law behavior $S_{\mathrm{R}}=\alpha\partial \mathrm{R}+S_{\mathrm{topo}}$ ($\partial \mathrm{R}$ is the length of the boundary of the region $\mathrm{R}$). The formula in Eq.~\eqref{KP} is designed in such a way that contributions from the area law term cancel.

The von Neumann entropy  $S_{\mathrm{R}}$ for a stabilizer Hamiltonian can be computed easily~\cite{linden_et_al:LIPIcs:2013:4327,PhysRevB.94.075151}. 
Let $G$ be the multiplicative group generated by all $\hat{A}_v$'s ($v\in\mathcal{V}$), $\hat{B}_p$'s ($p\in\mathcal{P}$), and possible closed string operators for which $|\Phi_0\rangle$ has the eigenvalue $+1$.  Suppose $|\Phi_0\rangle$ is the unique state that has the eigenvalue $+1$ for all operators in $G$. Then the projector onto $|\Phi_0\rangle$ can be written as
\begin{align}
|\Phi_0\rangle\langle\Phi_0|=\hat{P}_G\coloneqq\frac{1}{|G|}\sum_{\hat{g}\in G}\hat{g}.\label{projector}
\end{align}
We have $\hat{g}\hat{P}_G=\hat{P}_G\hat{g}=\hat{P}_G$ for any $\hat{g}\in G$ due to the rearrangement theorem.
To see Eq.~\eqref{projector}, it is enough to check that $\hat{P}_G|\Phi_0\rangle=|\Phi_0\rangle$ and $\hat{P}_G|\Psi\rangle=0$ if there exists $\hat{g}_*\in G$ such that $\hat{g}_*|\Psi\rangle=\lambda_*|\Psi\rangle$ with $\lambda_*\neq1$. The former is simply the definition of $|\Phi_0\rangle$. The latter follows by applying
$\hat{P}_G=\hat{P}_G\hat{g}_*$ to the state $|\Psi\rangle$:
\begin{align}
(\hat{P}_G|\Psi\rangle)=\hat{P}_G\hat{g}_*|\Psi\rangle=\lambda_*(\hat{P}_G|\Psi\rangle).
\end{align}

\begin{figure}
\includegraphics[width=0.5\textwidth]{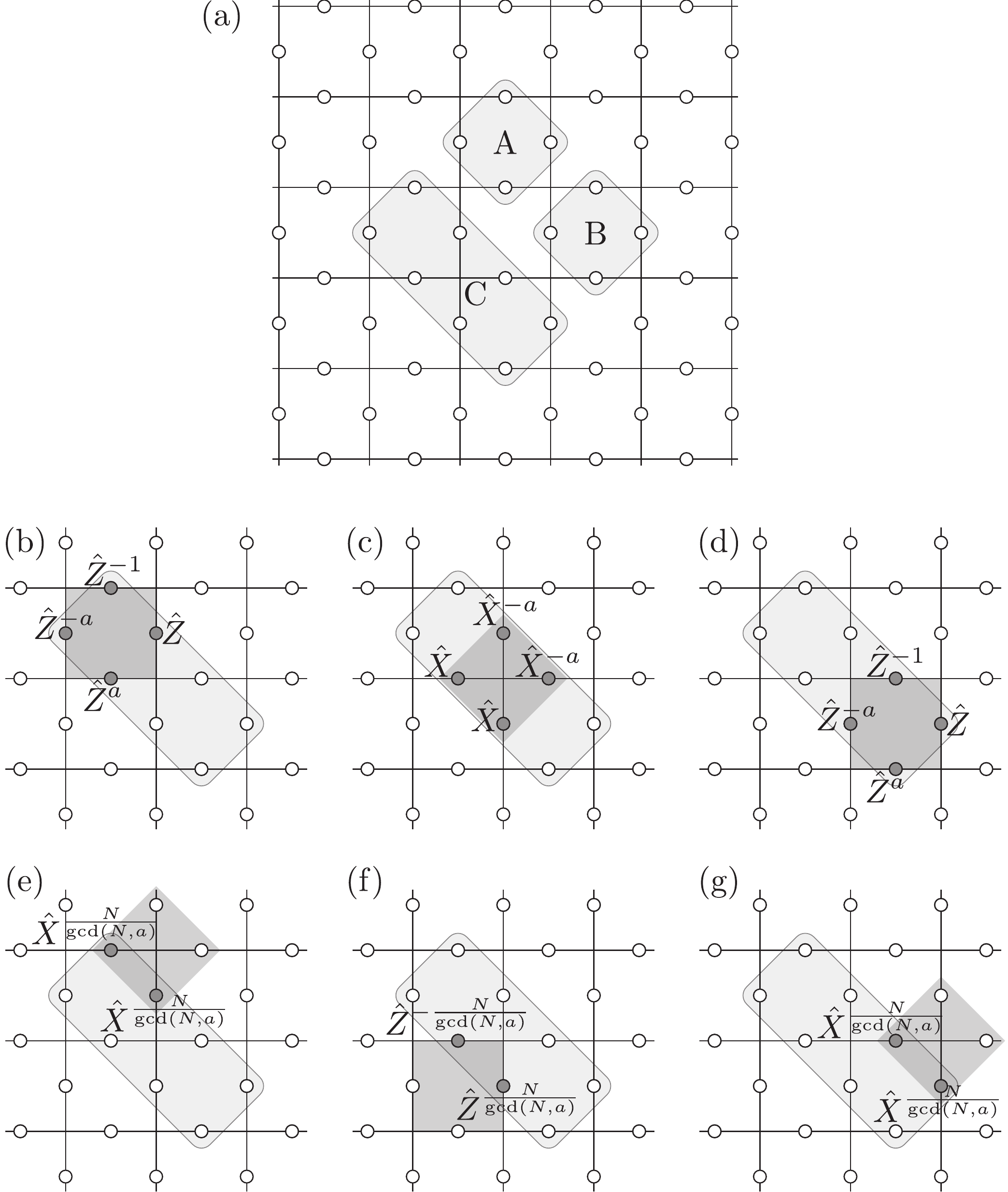}
\caption{\label{fig2} 
(a) Subregions A, B, C used in the computation of $S_{\mathrm{topo}}$.
(b)--(g): Generators of $G_{\mathrm{C}}$. (b)--(d) represent $\hat{A}_v$ and $\hat{B}_p$ themselves, and (e)--(g) correspond to $\hat{A}_v^{\frac{N}{\mathrm{gcd}(N,a)}}$ and $\hat{B}_p^{\frac{N}{\mathrm{gcd}(N,a)}}$.
When $N=N_1N_2$ and $a=N_1a'$ ($N_1$, $N_2$, and $a'$ are mutually coprime), then  $\frac{N}{\mathrm{gcd}(N,a)}=N_2$.
}
\end{figure}

As $\mathrm{tr}[\hat{g}]$ is nonzero only when $\hat{g}$ is identity, the order of the group $G$ is given by $|G|=N^{2L_1L_2}$. Similarly, $\mathrm{tr}_{\bar{\mathrm{R}}}[\hat{g}]$ can be nonzero only when $\hat{g}$ is identity over $\bar{\mathrm{R}}$. Thus
\begin{align}
\hat{\rho}_{\mathrm{R}}=\frac{1}{|G|}\sum_{\hat{g}\in G}\mathrm{tr}_{\bar{\mathrm{R}}}[\hat{g}]=\frac{1}{N^{n_{\mathrm{R}}}}\sum_{\hat{g}\in G_{\mathrm{R}}}\hat{g}=\frac{|G_{\mathrm{R}}|}{N^{n_{\mathrm{R}}}}\hat{P}_{G_{\mathrm{R}}},
\end{align}
where $n_{\mathrm{R}}$ is the number of $N$-level spins in $\mathrm{R}$ and $G_{\mathrm{R}}$ is the subgroup of $G$ supported in $\mathrm{R}$. In the last step, we introduced the projector
\begin{align}
\hat{P}_{G_{\mathrm{R}}}\coloneqq\frac{1}{|G_{\mathrm{R}}|}\sum_{\hat{g}\in G_{\mathrm{R}}}\hat{g}.
\end{align}
Therefore, $\hat{\rho}_{\mathrm{R}}$ has only one nonzero eigenvalue $\lambda=|G_{\mathrm{R}}|/N^{n_{\mathrm{R}}}$, whose order of degeneracy is $n_\lambda=N^{n_{\mathrm{R}}}/|G_{\mathrm{R}}|=1/\lambda$. Therefore,\cite{linden_et_al:LIPIcs:2013:4327,PhysRevB.94.075151}
\begin{align}
&S_{\mathrm{R}}=-n_\lambda\lambda \log\lambda=n_{\mathrm{R}}\log N-\log |G_{\mathrm{R}}|.\label{TEE1}
\end{align}
Up to this point, no assumption has been made on $a$.

When $a$ is coprime to $N$, $|G_{\mathrm{R}}|$ is given by $N^{m_{\mathrm{R}}}$, where $m_{\mathrm{R}}$ is the number of generators of $G$ supported in $\mathrm{R}$.\cite{PhysRevB.94.075151}
Therefore, the formula in Eq.~\eqref{TEE1} reduces to
\begin{align}
S_{\mathrm{R}}=(n_{\mathrm{R}}-m_{\mathrm{R}})\log N.
\end{align}
Using this formula, we find that the topological entanglement entropy of our model is
\begin{align}
S_{\mathrm{topo}}=-\log N,\label{TEEresult}
\end{align}
regardless of $L_1$ and $L_2$, as far as $a$ is coprime to $N$. For example, for the subregions A, B, and C illustrated in Fig.~\ref{fig2} (a), we have
\begin{align}
\frac{S_{\mathrm{topo}}}{\log N}=(3+3+5)-(5+7+7)+7=-1.
\end{align}
We confirm this result by the exact diagonalization up to $L_1=L_2=3$ and $N=3$.

When $N$ and $a$ have a common divisor, one needs to directly use the formula in Eq.~\eqref{TEE1}. 
For example, let us take positive, mutually coprime integers $N_1$, $N_2$, $a'$ and set $N=N_1N_2$ and $a=N_1a'$.
For the subregions A, B, and C illustrated in Fig.~\ref{fig2} (a), we find
\begin{align}
&S_{\mathrm{topo}}\notag\\
&=[(3\log N-\log N_1)+(3\log N-\log N_1)+(5\log N-3\log N_1)]\notag\\
&\quad-[(5\log N-3\log N_1)+(7\log N-3\log N_1)+(7\log N-3\log N_1)]\notag\\
&\quad+(7\log N-3\log N_1)\notag\\
&=-\log N+\log N_1=-\log N_2=-\log N_a.
\end{align}
Generators of $G_{\mathrm{R}}$ used in the calculation are shown in Fig.~\ref{fig2} (b)--(g) using the region C as an example.
This result is what one would expect from the $\mathbb{Z}_{N_a}$ topological order.
However, more generally, we have
\begin{align}
S_{\mathrm{topo}}&=-\log N+\log[\mathrm{gcd}(N,a)]\notag\\
&=-\log N_a-\log\Big[\frac{N}{N_a\mathrm{gcd}(N,a)}\Big]\label{generalStopo}.
\end{align}
By definition (see Eqs.~\eqref{Npf} and \eqref{Na}), $N/[N_a\mathrm{gcd}(N,a)]$ is a positive integer. When it is larger than one, $S_{\mathrm{topo}}$ is shifted from the expected value $-\log N_a$. We examine this additional contribution to $S_{\mathrm{topo}}$ in detail below.

\begin{figure}
\includegraphics[width=0.5\textwidth]{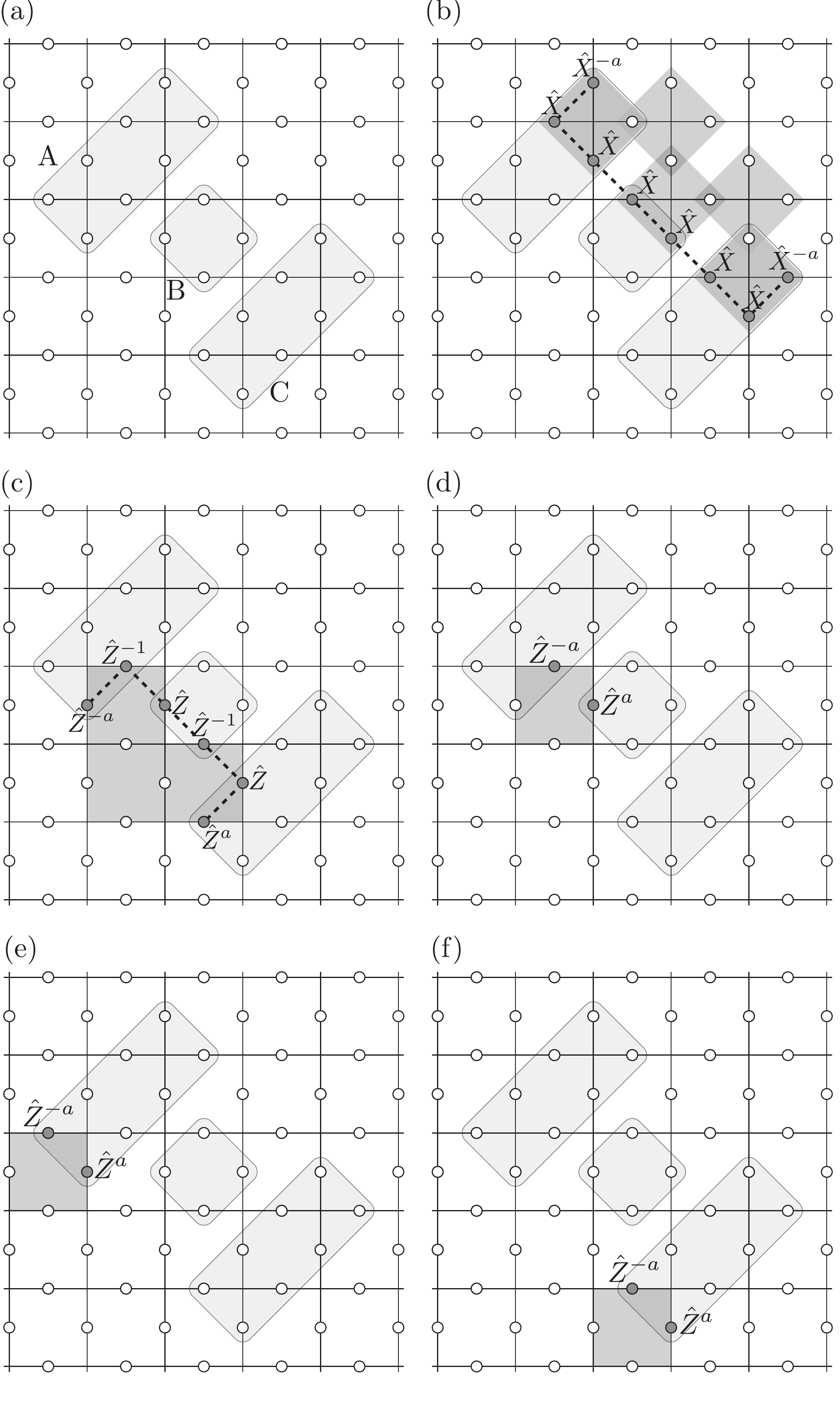}
\caption{\label{fig22} 
(a) Subregions A, B, C used in the computation of $S_{\mathrm{dumb}}$.
(b)--(g): Generators of $G_{\mathrm{ABC}}$. Those simply given by $\hat{A}_v$ and $\hat{B}_p$ are omitted.
(b),(c) correspond to subsystem symmetry operators. 
(d)--(f) correspond to $\hat{B}_p^a$.
}
\end{figure}

\subsection{Spurious contributions}
It is known that the topological entanglement entropy may suffer from spurious contributions and may become nonzero even when the ground state does not have a topological order~\cite{PhysRevB.94.075151,PhysRevLett.122.140506,PhysRevB.100.115112}. Thus we need to verify that the nonzero topological entanglement entropy found in the previous section is the legitimate one.

In Ref.~\onlinecite{PhysRevLett.122.140506}, it was shown that such spurious contributions can be captured by another combination of entropies computed for a dumbbell shape configuration:
\begin{align}
S_{\mathrm{dumb}}\coloneqq(S_{\mathrm{ABC}}+S_{\mathrm{B}})-(S_{\mathrm{AB}}+S_{\mathrm{BC}}).
\end{align}
Regions A, B, C must be chosen carefully~\cite{PhysRevLett.122.140506}, and here we assume those illustrated in Fig.~\ref{fig22}(a).
 
When $a$ is coprime to $N$, we find
\begin{align}
&\frac{S_{\mathrm{dumb}}}{\log N}=(11+3)-(7+7)=0,
\end{align}
implying that $S_{\mathrm{topo}}$ in Eq.~\eqref{TEEresult} is physical. This remains true more generally when $N/[N_a\mathrm{gcd}(N,a)]=1$.

This is no longer the case when $N/[N_a\mathrm{gcd}(N,a)]>1$.
For example, when $N=a^2$ ($a>1$),
there are no anyons ($N_a=1$) and the phase must be topologically trivial as we will discuss in Sec.~\ref{sec:case2}. However, in this case, $N/[N_a\mathrm{gcd}(N,a)]=a>1$ and $S_{\mathrm{topo}}$ in Eq.~\eqref{generalStopo} 
becomes
\begin{align}
S_{\mathrm{topo}}=0-\log a=-0.5\log N.\label{Stopo2}
\end{align}
This nonzero value comes from the spurious contribution originating from subsystem symmetries.
Subsystem symmetries are rigid string operators that cannot be deformed freely, unlike the Wilson loop operators, but commute with the Hamiltonian.
In our model, subsystem symmetries exist when $N_a\neq N$. They have nontrivial contribution to $S_{\mathrm{topo}}$ and $S_{\mathrm{dumb}}$ when their ends have a shape illustrated by dashed lines in Fig.~\ref{fig2} (b),(c), which occurs when $N/[N_a\mathrm{gcd}(N,a)]>1$. Indeed, when $N=a^2$ ($a>1$), we find
\begin{align}
&\frac{S_{\mathrm{dumb}}}{\log N}=(7.5+2.5)-(5.5+5.5)=-1.\label{Sdumb2}
\end{align}
We illustrate generators of $G_{\mathrm{ABC}}$ used in the calculation in Fig.~\ref{fig22}(b)--(f).
These behaviors imply that $N=a^2$ cases realize subsystem symmetry-protected topological (SSPT) phases and we will come back to this point in Sec.~\ref{sec:sspt}.

\subsection{Anyons}
\label{sec:anyons}

When $a$ is coprime to $N$, all magnetic and electric excitations can be understood as anyons with nontrivial mutual braiding statistics. They are created in pairs by open string operators as we saw in Sec.~\ref{local}, or by extended string operators in Eqs.~\eqref{Xp1} and ~\eqref{Zv1} without forming a pair. 
 The appearance of anyonic excitations is another hallmark of topologically ordered phases.

When $a$ is not coprime to $N$, some of magnetic and electric excitations are trivial in the sense they can be created locally without forming a pair. To see this,
let us focus on divisors of $N$ given by
\begin{align}
d_k\coloneqq \frac{N}{\mathrm{gcd}(a^k,N)}\in D_N\quad(k=1,2,3,\cdots).
\end{align}
If $k<k'$, $d_k/d_{k'}$ is a positive integer, because 
\begin{align}
\frac{d_k}{d_{k'}}=\frac{\mathrm{gcd}(a^{k'},N)}{\mathrm{gcd}(a^k,N)}=\mathrm{gcd}\Big(\frac{a^{k'}}{\mathrm{gcd}(a^k,N)},\frac{N}{\mathrm{gcd}(a^k,N)}\Big)\geq1.
\end{align}
In particular, $d_k=N_a$ for every $k\geq \mathrm{max}\{r_j\}_{j=m+1}^n$, where $N_a$ was defined in Eq.~\eqref{Na} and $r_j$'s are powers appearing the prime factorization in Eq.~\eqref{Npf}. Therefore, all $d_k$'s are multiples of $N_a$.

The string operator
\begin{align}
\hat{X}_{v,d_{k}}^{(1)}\coloneqq[\hat{X}_{v,v+(k,0)}^{(1)}]^{d_{k}}&=\prod_{\ell=0}^{k-1}\hat{X}_{(m_1+1+\ell,m_2+\frac{1}{2})}^{d_{k}a^\ell}
\end{align}
creates a single magnetic excitation with the eigenvalue $\omega^{d_{k}}$ of $\hat{B}_{v}$. The eigenvalue of $\hat{B}_{v+(k,0)}$ remains $\omega^{-d_{k}a^{k}}=1$.
We can do the same for electric excitations.
Hence, a magnetic or electric excitation with the eigenvalue $\omega^{\ell N_a}$ ($\ell\in\mathbb{Z}$) can be created locally by $[\hat{X}_{v,N_a}^{(1)}]^\ell$ without forming a pair. Conversely, if $q$ is not a multiple of $N_a$, excitations with eigenvalue $\omega^q$ needs to be created in pairs. Therefore, only excitations with the charge $q=1,2,\cdots,N_a-1$ are nontrivial.

Generally, we label the anyonic excitations by their electric and magnetic charges $q_e$ and $q_m$, where $q_e, q_m\in \{0,1,\cdots, N_a-1\}$. 
The topological order of this model is thus identical to that of the standard $\mathbb{Z}_{N_a}$ toric code model, i.e. the same anyon types, fusion rules and braiding statistics.  In particular, they satisfy the following fusion rule:
\begin{equation}
  (q_e,q_m)\times (q_e', q_m')=([q_e+q_e']_{N_a}, [q_m+q_m']_{N_a}).
\end{equation}
Here $[x]_{N_a}$ means $x$ mod $N_a$. Thus we may view the anyons as an Abelian group $\mathcal{A}=\mathbb{Z}_{N_a}\times \mathbb{Z}_{N_a}$, with the multiplication given by fusion.

However, if we take into account lattice translation symmetry, the system can have distinct translation symmetry-enriched topological phases~\cite{SET} as the standard $\mathbb{Z}_{N_a}$ toric code. More specifically,
Under a unit translation in $x_1$ or $x_2$, an anyon $(q_e, q_m)$ becomes
\begin{equation}
	T_i:(q_e, q_m)\rightarrow (aq_e, a^{-1}q_m).
	\label{}
\end{equation}
This action is well-defined, since for every $q=1,2,\cdots,N_a-1$, there exists $\ell$ ($1\leq \ell\leq N_a-1$) such that $q=a^\ell$ mod $N_a$. Then $aq\coloneqq a^{\ell+1}$ and $a^{-1}q\coloneqq a^{\ell-1}$ mod $N_a$. When $q=0$, $aq=a^{-1}q=0$.

We should mention that to completely describe the symmetry-enriched topological order there are further information beyond the permutation action~\cite{SET}. But they are not relevant for our purpose, so we will not consider them in more details.

\subsection{Symmetry defects}
\label{sec:defect}

When $a\neq 1$, the $T_i$ action generally changes anyon types. We can also see that $T_i^{M_{N_a}(a)}$ keeps all anyon types invariant, so effectively $T_i$ generates a $\mathbb{Z}_{M_{N_a}(a)}$ symmetry group of the low-energy topological theory.  In this section we will use $\rho_{a^k}$ to denote the permutation
\begin{equation}
  \rho_{a^k}: (q_e, q_m)\rightarrow (a^kq_e, a^{-k}q_m).
\end{equation}

Before we continue, it will be very useful to understand the properties of (point-like) symmetry defects i.e. dislocations in this case~\cite{Bombin:2010xn, You:2012sfg,Barkeshli:2013yta, Teo2014, SET, Teo2015,TarantinoSET2016}.  Generally, each symmetry defect is uniquely associated with a group element, which determines the symmetry action that takes place when moving around the defect. We denote the set of all defects associated with symmetry group element $g$ by $\mathcal{C}_g$. Note that for $g=1$, trivial defects are nothing but the anyons. Symmetry defects are always at the end points of defect lines, which can be intuitively thought of as branch cuts where the symmetry action takes place. Just like anyons, defects can fuse with each other to new defects, and the fusion rules must respect the group multiplication structure. Defects can also fuse with anyons, which do not change the associated group element. See Ref.~\onlinecite{SET} for a more systematic discussion of defect fusion rules.

Let us consider the $\rho_{a^k}$ defects. We pick one of them as a reference and denote it by $\sigma_{a^k,0}$. The other defects can be obtained by fusing $\sigma_{a^k,0}$ with anyons. Naively, one might think that the number of different defect types is the same as the number of anyon types. However, due to the permutation action, we also have the following fusion rule:
\begin{equation}
  ((a^k-1)q_e, (a^{-k}-1)q_m)\times \sigma_{a^k,0}=\sigma_{a^k,0},
\end{equation}
for any $q_e, q_m$.  To see this, one can locally create a pair of anyons $(q_e,q_m)$ and $(-q_e,-q_m)$ near the defect, move $(q_e, q_m)$ around the defect so it becomes $(a^kq_e, a^{-k}q_m)$, and then fuse it again with $(-q_e,-q_m)$ to give $((a^k-1)q_e, (a^{-k}-1)q_m)$. In other words, $\sigma_{a^k,0}$ and $\sigma_{a^k,0}\times ((a^k-1)q_e, (a^{-k}-1)q_m)$ are related by a local operation, so must be the same type of defect.

Therefore, the defect types should be identified with a quotient of the group of anyons $\cal A$ by the subgroup generated by $(a^k-1,0)$ and $(0, a^{-k}-1)$~\cite{SET, Teo2015}. We will denote by $[q_e,q_m]$ the equivalence classes of anyons under this quotient. Define $t_{a^k}=\gcd(a^k-1, N_a)=\gcd(a^{-k}-1, N_a)$ (the second equality follows from $\gcd(a^k, N_a)=1$), then we can label the defects by $\sigma_{a^k,[q_e, q_m]}$ where $q_{e,m}=0,1,\cdots t_{a^k}$ as representatives of the equivalence classes:
\begin{equation}
(q_e, q_m)\times \sigma_{a^k,0}=\sigma_{a^k,[q_e,q_m]}.
\end{equation}
These different types of defects can be uniquely labeled by the braiding phases of  $\rho_{a^k}$-invariant anyons around the defect. We can now define $\sigma_{a^k,0}$ as the defect where all such braiding phases are $1$. 

As an example, if $N_a$ is a prime and $a\neq 1$ mod $N_a$, then the subgroup generated by $a^k-1$ for $0<k< N_a-1$ is basically the entire group $\mathbb{Z}_{N_a}$. So the quotient group has a single element and there is only a unique type of defect.

We also need to know how the $\rho_{a^k}$ defects transform under the $\rho_{a^{k'}}$ action. It is clear that $\sigma_{a^k,0}$ is invariant under $\rho_{a^{k'}}$. So the action on $\sigma_{a^k, [q_e, q_m]}$ is given by 
\begin{equation}
  \rho_{a^{k'}}: \sigma_{a^k, [q_e, q_m]}\rightarrow \sigma_{a^k, [a^{k'}q_e, a^{-k'}q_m]}.
  \label{rhodefect}
\end{equation}

Let us now consider the ground state degeneracy on a torus, with a $\rho_{a^{k_1}}$ defect line in one direction and a $\rho_{a^{k_2}}$ defect line in the other direction. According to the general theory in Ref.~\onlinecite{SET}, the ground state degeneracy is equal to the number of $\rho_{a^k}$ defect types invariant under $\rho_{a^{k'}}$ action given in Eq.~\eqref{rhodefect}. 

We now show that the number of such $\rho_{a^{k_1}}$ defects is 
\begin{equation}
  \gcd(a^k-1, a^{k'}-1, N_a)^2.
\end{equation}
To see why,  first notice that the invariance of $\sigma_{a^k, [q_e,q_m]}$ under $\rho_{a^{k'}}$ means that  $q_e, q_m$ satisfy 
\begin{equation}
  [(a^{k'}-1)q_e, (a^{-k'}-1)q_m]=[0,0].
  \label{defect-inv-eq}
\end{equation}
Without any loss of generality, we can restrict $q_e,q_m\in \{0,1,\cdots, t_{a^{k}}\}$.
Clearly we can treat the electric and magnetic sector separately, so we will focus on the electric sector and suppress the subscript $e$. To shorten notations,
define $b_1=a^{k}-1, b_2=a^{k'}-1$, and $t_i = \gcd(b_i, N_a)$. In the electric sector, Eq.~\eqref{defect-inv-eq} means that there exists an integer $r$ such that 
\begin{equation}
  b_2q \equiv b_1r\text{ mod }N_a.
\end{equation}
 Given a $q$, this is possible if and only if $t_1=\gcd(b_1, N_a)$ divides $b_2q$. In other words, there exists an integer $r'$ such that 
 \begin{equation}
  b_2q=t_1r'.
 \end{equation}
  The smallest positive integer $q$ that makes it solvable is $\frac{t_1}{\gcd(t_1, b_2)}$. Note that $\gcd(t_1,b_2)=\gcd(\gcd(b_1,N_a),b_2)=\gcd(b_1,b_2, N_a)$. Therefore the number of solution is precisely $\gcd(b_1,b_2,N_a)$. The same argument works for the magnetic sector, so together we find the total number of solutions to Eq.~\eqref{defect-inv-eq} is given by $\gcd(b_1,b_2,N_a)^2=\gcd(a^k-1, a^{k'}-1, N_a)^2$.

We now show that knowing the permutation action of $T_i$ on anyons is enough to derive the topological degeneracy. Here the key is to think of a $L_1\times L_2$ torus as a torus in continuum, but with a $T_1^{L_1}$ defect line along $x_2$, and $T_2^{L_2}$ defect line along $x_1$. Intuitively this is because traveling across the torus in the $x_i$ direction is the same as translating by $L_i$. With this picture, the ground state degeneracy is obtained by substituing $k=L_1$ and $k'=L_2$, which reproduces the result in Eq.~\eqref{ndeggeneral}. In our model, as shown in Sec. \ref{sec:relation} when $a$ and $N$ are coprime we can indeed explicitly map the Hamiltonian on a torus to the standard toric code (where translation symmetry acts trivially) with twisted boundary condition, or equivalently with symmetry defect lines wrapping around the two non-contractible cycles, in full agreement with the argument in this section. The standard toric code has a smooth continuum limit, thus the finite-lattice effect is completely captured by the defect lines, establishing the continuum picture at the microscopic level.

\section{Phases with no topological order}
\label{sec:case2}
In this section, we consider the case when $a$ is a multiple of $\mathrm{rad}(N)$.

\subsection{Uniqueness of the ground state}
Let us demonstrate the uniqueness of the ground state regardless of the choice of the system size $L_1$ and $L_2$ although it is already implied by our general formula in Eq.~\eqref{ndeggeneral} with $N_a=1$.

Let $\ell_0$ be the smallest positive integer such that 
\begin{align}
a^{\ell_0}=0\mod N.
\end{align}
To see that $\ell_0$ indeed exists, let us write $r_{\mathrm{M}}\coloneqq \mathrm{max}\{r_i\}_{i=1}^{n}$, where $r_i$'s are powers appearing in the prime factorization of $N$ in  Eq.~\eqref{Npf}. Because $\mathrm{rad}(N)^{r_{\mathrm{M}}}=\prod_{j=1}^{n}p_j^{r_{\mathrm{M}}}$
is a multiple of $N$, and also because $a^{r_{\mathrm{M}}}$ is a multiple of $\mathrm{rad}(N)^{r_{\mathrm{M}}}$, we have
\begin{align}
a^{r_{\mathrm{M}}}=0\mod N.
\end{align}
Therefore, $\ell_0$ is in the range $1\leq \ell_0\leq r_{\mathrm{M}}$.

Then, our discussion in Sec.~\ref{sec:gs} implies that the state
\begin{align}
\hat{X}_{(m_1+\frac{1}{2},m_2+\frac{1}{2}),(m_1+\ell_0+\frac{1}{2},m_2+\frac{1}{2})}^{(1)}|\Phi_0\rangle\label{string3}
\end{align}
contains a magnetic excitation with eigenvalue $\omega$ at the plaquette $p=(m_1+\frac{1}{2},m_2+\frac{1}{2})$. The eigenvalue of the plaquette operator $\hat{B}_{(m_1+\ell_0+\frac{1}{2},m_2+\frac{1}{2})}$ remains $1$. Most importantly, the string operator in Eq.~\eqref{string3} is local in the sense its length $\ell_0$ does not depend on the system size.
Hence, a \emph{single elementally} magnetic excitation can be created locally. Similarly, the state
\begin{align}
\hat{Z}_{(m_1-\ell_0,m_2),(m_1,m_2)}^{(1)}|\Phi_0\rangle
\end{align}
contains an electric excitation with the eigenvalues $\omega^{-1}$ at the vertex $v=(m_1,m_2)$.  
The rest of the discussion proceeds exactly the same as in Sec.~\ref{sec:unique}. Therefore,
\begin{align}
N_{\mathrm{deg}}=1
\end{align}
for any $L_1$ and $L_2$. 

\subsection{Example 1: $N=a$}
As an example, let us discuss the case of $N=a$.
In this case, the Hamiltonian is completely decoupled:
\begin{align}
\hat{H}&=\sum_{\bm{r}\in\Lambda}\hat{h}_{\bm{r}},\\
\hat{h}_{(m_1,m_2)}&=\frac{1}{2}\Big(\hat{X}_{(m_1-\frac{1}{2},m_2)}\hat{X}_{(m_1,m_2-\frac{1}{2})}+\text{h.c.}\Big)\notag\\
&\quad-\frac{1}{2}\Big(\hat{Z}_{(m_1,m_2-\frac{1}{2})}\hat{Z}_{(m_1-\frac{1}{2},m_2)}^{-1}+\text{h.c.}\Big).
\end{align}
The ground state of $\hat{h}_{\bm{r}}$ is unique and has the energy gap $\Delta_1$. We denote the ground state by $|\phi_0\rangle_{\bm{r}}$. Then the unique ground state of $\hat{H}$ is given by the product state $\otimes_{\bm{r}\in\Lambda}|\phi_0\rangle_{\bm{r}}$. Therefore, this phase is completely trivial. Indeed, the topological entablement entropy in Eq.~\eqref{KP} vanishes
\begin{align}
\frac{S_{\mathrm{topo}}}{\log N}=(3+3+5)-(5+8+8)+10=0.\label{Stopo3}
\end{align}
for the subregions in Fig.~\ref{fig2}(a).

\begin{figure*}
\includegraphics[width=\textwidth]{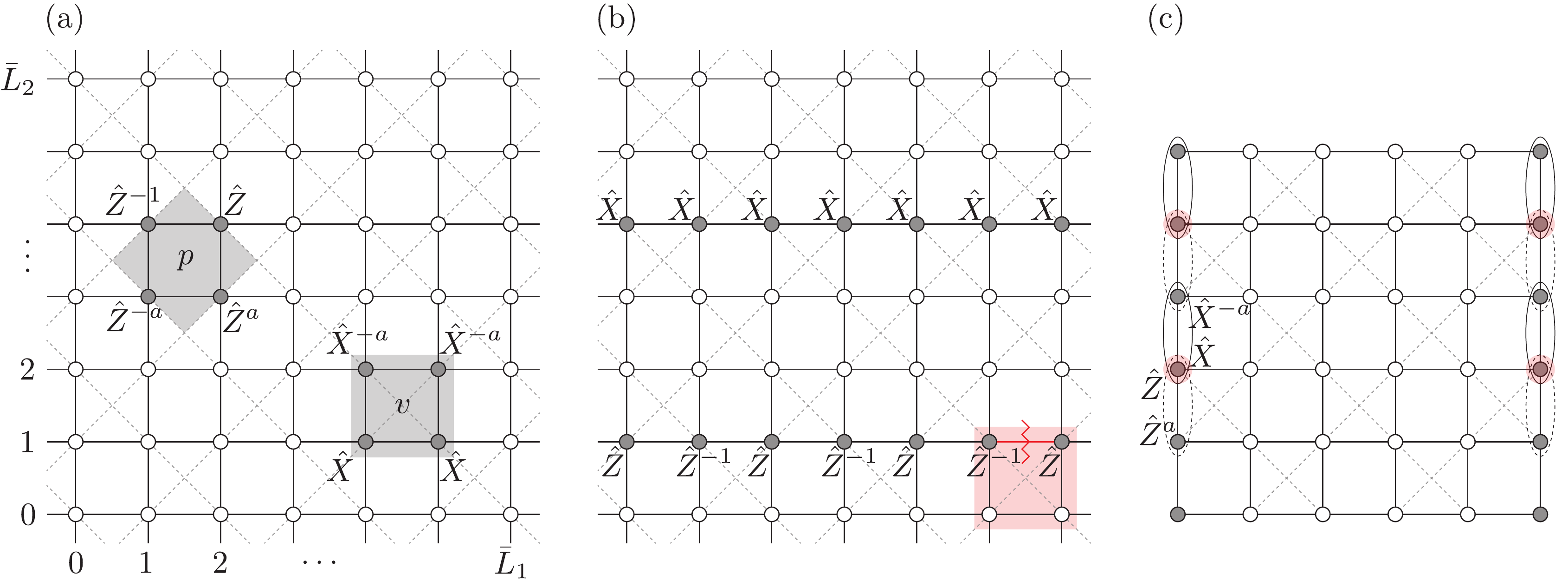}
\caption{\label{fig3} 
The model rotated by 45 degree, corresponding to the $\bar{L}_1=\bar{L}_2=6$ case. Dashed lines represent the original square lattice before the rotation.
(a) Plaquette operators and vertex operators. 
(b) Subsystem symmetries.  The red jagged line represents the symmetry flux across the link between $(\bar{L}_1-1,2j_2+1)$ and $(0,2j_2+1)$.
(c) Edge zero modes under open boundary condition protected by subsystem symmetries.
}
\end{figure*}

\subsection{Example 2: $N=a^2$}
\label{sec:sspt}
Next we discuss the case of $N=a^2$. We argue that this case realizes a SSPT phase\cite{PhysRevB.98.035112,PhysRevB.98.235121}.

To this end, we study the property of the model obtained by rotating the one introduced in Sec.~\ref{sec:model} by $45$ degree (see Fig.~\ref{fig3}).
Spins are now defined on square lattice sites $\bm{r}=(\bar{m}_1,\bar{m}_2)$ with $\bar{m}_1,\bar{m}_2\in\mathbb{Z}$. Vertices and plaquettes can be associated with odd (even) sites:
\begin{align}
&\bar{\mathcal{V}}\coloneqq\{(\bar{m}_1,\bar{m}_2)\,|\,(-1)^{\bar{m}_1+\bar{m}_2}=-1\},\\
&\bar{\mathcal{P}}\coloneqq\{(\bar{m}_1,\bar{m}_2)\,|\,(-1)^{\bar{m}_1+\bar{m}_2}=+1\}.
\end{align}
The Hamiltonian is given by
\begin{align}
\hat{H}\coloneqq-\sum_{v\in \bar{\mathcal{V}}}\frac{1}{2}(\hat{A}_v+\text{h.c.})-\sum_{p\in \bar{\mathcal{P}}}\frac{1}{2}(\hat{B}_p+\text{h.c.}),
\end{align}
where
\begin{align}
\hat{A}_{(\bar{m}_1,\bar{m}_2)}&=\hat{X}_{(\bar{m}_1+1,\bar{m}_2+1)}^{-a}\hat{X}_{(\bar{m}_1,\bar{m}_2+1)}^{-a}\hat{X}_{(\bar{m}_1,\bar{m}_2)}\hat{X}_{(\bar{m}_1+1,\bar{m}_2)},\label{Av3}\\
\hat{B}_{(\bar{m}_1,\bar{m}_2)}&=\hat{Z}_{(\bar{m}_1+1,\bar{m}_2+1)}\hat{Z}_{(\bar{m}_1,\bar{m}_2+1)}^{-1}\hat{Z}_{(\bar{m}_1,\bar{m}_2)}^{-a}\hat{Z}_{(\bar{m}_1+1,\bar{m}_2)}^{a}.\label{Bp3}
\end{align}

\subsubsection{Charge pumping}
Let us work with the periodic boundary condition first. We identify $\bm{r}+(n_1\bar{L}_1,n_2\bar{L}_2)$ with $\bm{r}$ for $n_1,n_2\in\mathbb{Z}$. Both $\bar{L}_1$ and $\bar{L}_2$ are assumed to be even. Unit translation symmetries of the model shift $\bm{r}$ either by $(1,1)$ or $(1,-1)$.

The model has subsystem symmetries 
\begin{align}
&\hat{X}_{\bar{m}_2}\coloneqq\prod_{\bar{m}_1=0}^{\bar{L}_1-1}\hat{X}_{(\bar{m}_1,\bar{m}_2)},\label{sss1}\\
&\hat{Z}_{\bar{m}_2}\coloneqq\prod_{\bar{m}_1=0}^{\bar{L}_1-1}\hat{Z}_{(\bar{m}_1,\bar{m}_2)}^{(-1)^{\bar{m}_1}}\label{sss2}
\end{align}
for each $\bar{m}_2$ separately, as illustrated in Fig.~\ref{fig3} (b), which act only on a single row. When $N=a^2$, these operators can be rewritten in terms of stabilizers as
\begin{align}
&\hat{X}_{\bar{m}_2}=\prod_{j_1=0}^{\bar{L}_1/2}\hat{A}_{(2j_1+1,\bar{m}_2)}\hat{A}_{(2j_1,\bar{m}_2+1)}^a,\label{Xm21}\\
&\hat{Z}_{\bar{m}_2}=\prod_{j_1=0}^{\bar{L}_1/2}\hat{B}_{(2j_1-1,\bar{m}_2-1)}\hat{B}_{(2j_1,\bar{m}_2-2)}^a
\end{align}
when $\bar{m}_2$ is even, and
\begin{align}
&\hat{X}_{\bar{m}_2}=\prod_{j_1=0}^{\bar{L}_1/2}\hat{A}_{(2j_1,\bar{m}_2)}\hat{A}_{(2j_1+1,\bar{m}_2+1)}^a,\label{Xm22}\\
&\hat{Z}_{\bar{m}_2}=\prod_{j_1=0}^{\bar{L}_1/2}\hat{B}_{(2j_1,\bar{m}_2-1)}^{-1}\hat{B}_{(2j_1+1,\bar{m}_2-2)}^{-a}
\end{align}
when $\bar{m}_2$ is odd. Hence, in the ground state $|\Phi_0\rangle$ where all vertex operators and plaquette operators take the value $+1$, we have
\begin{align}
\hat{X}_{\bar{m}_2}|\Phi_0\rangle=\hat{Z}_{\bar{m}_2}|\Phi_0\rangle=|\Phi_0\rangle
\end{align}
for all $\bar{m}_2$.

Now we insert a symmetry flux associated with the subsystem symmetry $\hat{Z}_{\bar{m}_2=2j_2+1}$ at the link between $\bar{m}_1=\bar{L}_1-1$ and $\bar{m}_1=0$. This operation multiplies a factor $\omega^{-a}$ to the vertex term $\hat{A}_{(\bar{L}_1-1,2j_2)}$ [the red shaded vertex in Fig.~\ref{fig3} (b)]:
\begin{align}
\hat{H}'= \hat{H}+\frac{1-\omega^{-a}}{2}\hat{A}_{(\bar{L}_1-1,2j_2)}+\text{h.c.}
\end{align}
In the ground state $|\Phi_0'\rangle$ of $\hat{H}'$, the eigenvalue of $\hat{A}_{(\bar{L}_1-1,2j_2)}$ is thus modified to $\omega^a$. Therefore, using Eqs.~\eqref{Xm21} and ~\eqref{Xm22}, we find
\begin{align}
&\hat{X}_{2j_2}|\Phi_0'\rangle=\omega^a |\Phi_0'\rangle.
\end{align}
Namely, the charge $\omega^a$ is pumped for the subsystem symmetry $\hat{X}_{2j_2}$ upon inserting the symmetry flux associated with the subsystem symmetry $\hat{Z}_{2j_2+1}$. This pumped charge is a topological invariants that distinguishes this phase from product states.

\subsubsection{Zero energy edge states}
Next let us consider the open boundary condition. We impose the subsystem symmetries $\hat{X}_{\bar{m}_2}$ and $\hat{Z}_{\bar{m}_2}$ in Eqs.~\eqref{sss1} and ~\eqref{sss2} for every $\bar{m}_2$, including the edges.

We introduce two sets of generalized Pauli matrices
\begin{align}
\hat{\sigma}_{2j,L}^x\coloneqq\hat{X}_{(0,2j)}\hat{X}_{(0,2j+1)}^{-a},\\
\hat{\sigma}_{2j,L}^z\coloneqq\hat{Z}_{(0,2j)}\hat{Z}_{(0,2j-1)}^{a},
\end{align}
and
\begin{align}
\hat{\sigma}_{2j,R}^x\coloneqq\hat{X}_{(\bar{L}_1-1,2j)}\hat{X}_{(0,2j+1)}^{-a},\\
\hat{\sigma}_{2j,R}^z\coloneqq\hat{Z}_{(\bar{L}_1-1,2j)}\hat{Z}_{(0,2j-1)}^{a},
\end{align}
which commute with all stabilizers in the bulk Hamiltonian.
They satisfy
\begin{align}
\hat{\sigma}_{2j,s}^z\hat{\sigma}_{2j',s'}^x=\omega^{\delta_{s,s'}\delta_{j,j'}} \hat{\sigma}_{2j',s'}^x\hat{\sigma}_{2j,s}^z
\end{align}
for $s,s'=L,R$ and $j,j'=1,2,\cdots,\frac{\bar{L}_2}{2}-1$. A pair of $\hat{\sigma}_{2j,s}^x$ and $\hat{\sigma}_{2j,s}^z$ generates a $\mathbb{Z}_N\times\mathbb{Z}_N$ symmetry, implying $N$-fold degeneracy, and there are $\bar{L}_2-2$ such pairs. This $N^{\bar{L}_2-2}$-fold degeneracy cannot be lifted by perturbations on the edges, as long as the subsystem symmetries are maintained. In contrast, the two edges at $\bar{m}_2=0$ and $\bar{m}_2=\bar{L}_2-1$ can be gapped by edge perturbations.

\section{Discussions}
\label{sec:discussion}
As a concluding remark, let us discuss implications of our example on the Lieb--Schultz--Mattis (LSM) type theorems~\cite{LiebSchultzMattis,AffleckLieb,OshikawaYamanakaAffleck,YamanakaOshikawaAffleck,Oshikawa,Hastings,Hastings2,NachtergaeleSims,WatanabePNAS,ChengPRX,latticehomotopy,DominicThorngrenLSM,Bachmann,OgataTachikawaTasaki,TasakiReview}, which formulate necessary conditions for the unique ground state with nonzero excitation gap under the periodic boundary condition. When one of these conditions are not satisfied, the appearance of ground state degeneracy or gapless excitations is guaranteed. The ground state degeneracy originates either from spontaneous symmetry breaking or topological degeneracy. Hence, a violation of LSM type conditions in symmetric and gapped phases can be used as a sufficient condition for a nontrivial topological order~\cite{Hastings2,WatanabePNAS}. 

There are a variety of such theorems applicable to quantum many-body systems in different settings. 
For example, in one dimension, an early version of LSM theorems for quantum spin chains with spin-rotation symmetry state that $S$ needs to be an integer in the presence of the time-reversal symmetry~\cite{LiebSchultzMattis,AffleckLieb}. More generally,  $S-m$ ($m$ is the magnetization per unit cell) must be an integer to realize a unique gapped ground state~\cite{OshikawaYamanakaAffleck}.  Similarly, in fermionic systems with $U(1)$ symmetry, the filling $\nu$ (the average number of fermions per unit cell) must be an integer~\cite{YamanakaOshikawaAffleck}. These results apply to any sequence of $L_1$. One can even start with the infinite system from the beginning~\cite{OgataTachikawaTasaki,TasakiReview}.

In contrast, there is usually a restriction on the choice of the sequence of $L_i$'s in higher dimensional extensions of these theorems. In the formulation, one usually starts with a finite size system with the length $L_i$ in $x_i$ direction ($i=1,\cdots,d$) and considers the limit $L_1,\cdots,L_d\rightarrow +\infty$. 
For example, for spin systems, the arguments in Refs.~\onlinecite{Hastings,NachtergaeleSims} are effective only when $L_2,\cdots, L_d$ are all odd. For particle systems, the discussions in Refs.~\onlinecite{Oshikawa,Bachmann} assume that $L_2,\cdots, L_d$ are coprime to $q$ when $\nu=p/q$.  There is a way to remove such a restriction by modifying the boundary condition to a tilted one~\cite{YuanOshikawa}, but this argument is not about the original periodic boundary condition. Namely, changing the boundary condition from the periodic one to the tilted one might affect the degeneracy or excitation gap.

As we demonstrated through an example, a topologically ordered phase may not show topological degeneracy on torus depending on the sequence of system size. Hence, even when all the LSM type conditions are fulfilled and the ground state is indeed unique in some sequences of the system size, it still might be the case that the ground state is actually topologically ordered.

\begin{acknowledgements}
H.W. thanks Hiroki Hamaguchi for informing us of the proof of Eq.~\eqref{conjecture2} in Appendix~\ref{app:proof}.
The work of H.W. is supported by JSPS KAKENHI Grant No.~JP20H01825 and JP21H01789. M.C. acknowledges support from NSF under award number DMR-1846109. 
The work of Y.F. is supported by JSPS KAKENHI Grant No.~JP20K14402 and JST CREST Grant No.~JPMJCR19T2.
H.W. acknowledges the hospitality and fruitful discussions at the Institute of Basic Science, Daejeon, Korea, during the week of Conference on Advances in The Physics of Topological and Correlated Matter.
\end{acknowledgements}

\appendix

\section{Proof of Eq.~\eqref{conjecture}}
\label{app:proof}
Here we demonstrate the validity of Eq.~\eqref{conjecture}. As stated in the main text, we set
\begin{align}
\ell_2'=r\coloneqq\frac{\alpha_1+b_1N}{\mathrm{gcd}(\alpha_1+b_1N,\alpha_2+b_2N)}\label{defr1}.
\end{align}
Using the properties of the greatest common divisor, we find
\begin{align}
d_a&=\mathrm{gcd}(\alpha_1,\alpha_2,N)=\mathrm{gcd}(\alpha_1+b_1N,\alpha_2+b_2N,N)
\end{align}
and
\begin{align}
d_{1,a}&=\mathrm{gcd}(\alpha_1,N)=\mathrm{gcd}(\alpha_1+b_1N,N)\notag\\
&=d_a\mathrm{gcd}\Big(r\frac{\mathrm{gcd}(\alpha_1+b_1N,\alpha_2+b_2N)}{d_a},\frac{N}{d_a}\Big)\notag\\
&=d_a\mathrm{gcd}\Big(r,\frac{N}{d_a}\Big).
\end{align}
In the last line, we used the fact that $N/d_a$ is coprime to $\mathrm{gcd}(\alpha_1+b_1N,\alpha_2+b_2N)/d_a$. Hence, Eq.~\eqref{conjecture} can be rewritten as
\begin{align}
\mathrm{gcd}\Big(\frac{r}{\mathrm{gcd}\big(r,\frac{N}{d_a}\big)},d_a\Big)=1.\label{conjecture2}
\end{align}

Below we prove the following statement: for any integer $N\geq 2$ and integers $\alpha_1$ and $\alpha_2$ in the range $1\leq \alpha_1,\alpha_2 \leq N-1$, there always exist integers $b_1$ and $b_2$ such that Eq.~\eqref{conjecture2} holds\footnote{The proof is due to Hiroki Hamaguchi.} . In particular, $b_2$ can be set $0$. Since this is trivially the case when $d_a=1$, in the following we assume $d_a\neq1$.  We introduce shorthands $\alpha_1'\coloneqq \alpha_1/d_a$, $\alpha_2'\coloneqq \alpha_2/d_a$, and $N'\coloneqq N/d_a$.

For an integer $m$ and a prime $p$, let us denote by $\nu_p(m)$ the largest nonnegative integer $\nu$ such that $p^\nu$ divides $m$. 
Suppose that $e_j\coloneqq\nu_{p_j}(d_a)\geq1$ for $j=1,2,\cdots,J$. In other words, $d_a$ can be prime-factorized as $d_a=\prod_{j=1}^Jp_j^{e_j}$. 
Then, Eq.~\eqref{conjecture2} holds if and only if
\begin{align}
\nu_{p_j}(r)\leq \nu_{p_j}(N')\label{conjecture3}
\end{align}
for all $j=1,2,\cdots,J$. Also, by definition,
\begin{align}
\nu_{p_j}(r)&=\nu_{p_j}\left(\frac{\alpha_1'+b_1N'}{\mathrm{gcd}(\alpha_1'+b_1N',\alpha_2'+b_2N')}\right)\notag\\
&\leq \nu_{p_j}(\alpha_1'+b_1N').
\end{align}
Therefore, if
\begin{align}
\nu_{p_j}(\alpha_1'+b_1N')\leq\nu_{p_j}(N') \label{conjecture4}
\end{align}
simultaneously for all $j=1,2,\cdots,J$, Eq.~\eqref{conjecture3} is fulfilled. In the following, we write $n_j\coloneqq \nu_{p_j}(N')$ and $m_j\coloneqq \nu_{p_j}(\alpha_1')$.

Let us derive the condition for Eq.~\eqref{conjecture4}. When $n_j\geq m_j$, we need
\begin{align}
\nu_{p_j}\Big(\frac{\alpha_1'}{p_j^{m_j}}+b_1\frac{N'}{p_j^{m_j}}\Big)\leq n_j-m_j
\end{align}
with $\alpha_1'/p_j^{m_j}\neq0$ mod $p_j$. In this case, we can set $b_1=0$ mod $p_j$. 
On the other hand, when $m_j>n_j$, we need
\begin{align}
\nu_{p_j}\Big(\frac{\alpha_1'}{p_j^{n_j}}+b_1\frac{N'}{p_j^{n_j}}\Big)=0
\end{align}
with $\alpha_1'/p_j^{n_j}=0$ mod $p_j$ and $N'/p_j^{n_j}\neq0$ mod $p_j$. In this case, we can set $b_1=1$ mod $p_j$.
After all, we found a condition of the form $b_1=x_j$ mod $p_j$ for each $j=1,2,\cdots,J$. The Chinese remainder theorem guarantees the existence $b_1$ in the range $0$ to $-1+\prod_{j=1}^Jp_j$ such that these conditions are simultaneously satisfied.

\section{Reduction of generalized Pauli matrices}

When $N_1$ and $N_2$ are coprime, the $N=N_1N_2$-level spin can be decomposed into the tensor product of $N_1$- and $N_2$-level spins.
To see this, let us write the matrices in Eqs.~\eqref{defX} and ~\eqref{defZ} as $X(N)$ and $Z(N)$, respectively.  We have
\begin{align}
&VX(N)^{N_1+N_2}V^\dagger=X(N_1)\otimes X(N_2),\label{rf1}\\
&VZ(N)V^\dagger=Z(N_1)\otimes Z(N_2),\label{rf2}
\end{align}
where $[V]_{i_1i_2,i}\coloneqq \delta_{i-1,\text{mod}(N_2(i_1-1)+N_1(i_2-1),N)}$ and $[Z(N_1)\otimes Z(N_2)]_{i_1i_2,j_1j_2}\coloneqq[Z(N_1)]_{i_1,j_1}[Z(N_2)]_{i_2,j_2}$ for $i_1=1,\cdots,N_1$, $i_2=1,\cdots,N_2$, and $i=1,\cdots,N$. These reduction formulas can be readily shown by using the representations in Eqs.~\eqref{defX} and ~\eqref{defZ}.

Let us discuss the implication of these relations. Suppose that $N=N_1N_2$ and $N_1$ and $N_2$ are coprime. We introduce another modified Hamiltonian
\begin{align}
\hat{H}''&\coloneqq-\sum_{v\in\mathcal{V}}\frac{1}{2}(\hat{A}_v^{N_1+N_2}+\text{h.c.})-\sum_{p\in\mathcal{P}}\frac{1}{2}(\hat{B}_p+\text{h.c.}),
\end{align}
where $\hat{A}_v$ and $\hat{B}_p$ are vertex and plaquette operators in Eqs.~\eqref{Av} and \eqref{Bp} with $a_1=a_2=a$.
The eigenstates of this Hamiltonian are also identical to those for $\hat{H}$ in Eq.~\eqref{defH} and the ground state degeneracy remains unchanged. 

Let $\hat{V}$ be the global unitary operator whose action on each $N$ level spin is given by the unitary matrix $V$ above. Using the reduction formulas, we find
\begin{align}
\hat{V}\hat{H}''\hat{V}^\dagger
&=-\sum_{v\in\mathcal{V}}\frac{1}{2}(\hat{A}_v(N_1)\otimes\hat{A}_v(N_2)+\text{h.c.})\notag\\
&\quad-\sum_{p\in\mathcal{P}}\frac{1}{2}(\hat{B}_p(N_1)\otimes\hat{B}_p(N_2)+\text{h.c.}).
\label{defH3}
\end{align}
where $\hat{A}_v(N_i)$ and $\hat{B}_p(N_i)$ $(i=1,2)$ are vertex and plaquette operators for $N_i$-level spins. 
Ground states have the eigenvalue $+1$ for all $\hat{A}_v(N_i)$'s and $\hat{B}_p(N_i)$'s. This result indicates that, if we denote the ground state degeneracy of the our model for $N$-level spin by $N_{\mathrm{deg}}(N)$, 
\begin{align}
N_{\mathrm{deg}}(N)=N_{\mathrm{deg}}(N_1)N_{\mathrm{deg}}(N_2).
\end{align}
Indeed, this is consistent with our result in Eq.~\eqref{ndeggeneral} because
\begin{align}
&\mathrm{gcd}(a^{L_1}-1,a^{L_2}-1,N_1N_2)\notag\\
&=\mathrm{gcd}(a^{L_1}-1,a^{L_2}-1,N_1)\mathrm{gcd}(a^{L_1}-1,a^{L_2}-1,N_2),
\end{align}
when $N_1$ and $N_2$ are coprime.
More generally, for the form of $N$ in Eq.~\eqref{Npf}, we have
\begin{align}
N_{\mathrm{deg}}(N)&=\prod_{j=1}^nN_{\mathrm{deg}}(p_j^{r_j}).
\end{align}
However, this decomposition alone is not sufficient to derive Eq.~\eqref{ndeggeneral}. One still has to compute $N_{\mathrm{deg}}(p_j^{r_j})$ and this requires an investigation which is almost as hard as what we did in this work.

\end{document}